%% file: arxivVer.tex
\documentclass[prd,aps,superscriptaddress,nofootinbib,twocolumn]{revtex4-2}

\newif\ifuseBibTex\useBibTexfalse

\usepackage[utf8]{inputenc}
\usepackage[english]{babel}
\usepackage{amsmath}
\usepackage{amsfonts}
\usepackage{amssymb}
\usepackage{float}
\usepackage{graphicx}

\graphicspath{{./figures/}}
\usepackage{dcolumn} 
\usepackage{textcomp} 
\usepackage{xcolor}
\usepackage[normalem]{ulem} 
\usepackage{multirow}

\newcounter{univBibCtr}

\def\prn#1{{\left(#1\right)}}

\def\sbrk#1{{\left[#1\right]}}
\def\abrk#1{{\langle#1\rangle}}

\def\bra#1{{\langle#1|}}

\def\dbydt#1{{\frac{d #1}{dt}}}

\def\cg(#1,#2)(#3,#4)(#5,#6){\bra{#1,#2,#3,#4}#5,#6\rangle}

\def\ts#1{{_{\mbox{\scriptsize #1}}}}

\def\threej(#1,#2)(#3,#4)(#5,#6){\begin{pmatrix}#1&#3&#5\\#2&#4&#6\end{pmatrix}}
\def\sixj(#1,#2,#3)(#4,#5,#6){\begin{Bmatrix}#1&#2&#3\\#4&#5&#6\end{Bmatrix}}
\def\ninej(#1,#2,#3)(#4,#5,#6)(#7,#8,#9){\begin{Bmatrix}#1&#2&#3\\#4&#5&#6\\#7&#8&#9\end{Bmatrix}}

\def\mb{\mathbf}
\def\bs{\boldsymbol}

\newcommand{\complexScalar}{\phi} 
\newcommand{\NDegVac}{N} 
\newcommand{\SNot}{S_0} 
\newcommand{\HMode}{S} 
\newcommand{\axion}{a} 
\newcommand{\SBScale}{f_\text{SB}} 
\newcommand{\sigWidth}{\Delta x}
\newcommand{\surfTens}{\sigma_\text{DW}}
\newcommand{\coupling}{f_\text{int}}
\newcommand{\cnstRatio}{\xi}
\newcommand{\sigDur}{\Delta t}
\newcommand{\expVel}{\bar{v}}
\newcommand{\evtRate}{r} 
\newcommand{\RateBound}{R}
\newcommand{\totTime}{T}
\newcommand{\effTime}{\tilde{T}}

\newcommand{\truePos}{\epsilon } 
\newcommand{\totConf}{C} 
\newcommand{\energyDensity}{\rho_\text{DW}}
\newcommand{\pseudoMag}{B} 
\newcommand{\normPMag}{\mathcal{B}_\text{p}} 
\newcommand{\normPMagSens}{\mathcal{B}_\text{p}'} 
\newcommand{\SToNSym}{\zeta}

\newcommand{\MaxStoN}{12.6}
\newcommand{\BackgroundDuration}{10.7}

\newcommand{\sfrac}[2]{{}^#1{\mskip -5mu/\mskip -3mu}_#2} 
\newcommand{\PvalThr}{0.001}
\newcommand{\AngThr}{3.5}
\newcommand{\oneColWidth}{\columnwidth}
\newcommand{\twoColWidth}{\textwidth}

\begin{document}

\title{Search for topological defect dark matter with a global network of optical magnetometers} 

\input{authorList} 

\date{\today}

\begin{abstract}
Ultralight bosons such as axion-like particles are viable candidates for dark matter. They can form stable, macroscopic field configurations in the form of topological defects that could concentrate the dark matter density into many distinct, compact spatial regions that are small compared to the galaxy but much larger than the Earth. Here, we report the results of a search for transient signals from axion-like particle domain walls with the Global Network of Optical Magnetometers for Exotic physics searches (GNOME). We search the data, consisting of correlated measurements from optical atomic magnetometers located in laboratories all over the world, for patterns of signals propagating through the network consistent with domain walls. The analysis of data from a continuous month-long operation of the GNOME finds no statistically significant signals, thus placing experimental constraints on such dark matter scenarios.
\newpage
\end{abstract}

\maketitle

The nature of  dark matter, an invisible substance comprising over 80\% of the mass of the universe~\cite{Ber05,Gor14}, is one of the most profound mysteries of modern physics. Although evidence for the existence of dark matter comes from its gravitational interactions, unraveling its nature likely requires observing non-gravitational interactions between dark matter and ordinary matter~\cite{safronova2018search}. One of the leading hypotheses is that dark matter consists of ultralight bosons such as axions~\cite{Pre83} or axion-like particles (ALPs)~\cite{graham2015experimental,Gra15,co2020predictions}. Axions and ALPs arise from spontaneous symmetry breaking at an unknown energy scale $\SBScale$, which, along with their mass $m_\axion$, determines many of their physical properties.

ALPs can manifest as stable, macroscopic field configurations in the form of topological defects~\cite{Vil85,Pos13,Der14} or composite objects bound together by self-interactions such as boson stars~\cite{Bra16,kimball2018searching}. Such ALP field configurations could concentrate the dark matter density into many distinct, compact spatial regions that are small compared to the galaxy but much larger than the Earth. In such scenarios, Earth-bound detectors would only be able to measure signals associated with dark matter interactions on occasions when the Earth passes through such a dark-matter object. It turns out that there is a wide range of parameter space, consistent with observations, for which such dark-matter objects can have the required size and abundance such that the characteristic time between encounters could be on the order of one year or less~\cite{Pos13,Der14,kimball2018searching}. This opens up the possibility of searches with terrestrial detectors. Here we present the results of such a search for ALP domain walls, a class of topological defects which can form between regions of space with different vacua of an ALP field~\cite{Vil85, Pos13}. We note that while some models suggest that axion domain walls cannot survive to the present epoch~\cite{sikivie1982axions, press1989dynamical, buschmann_early-universe_2020}, there do exist a number of ALP models demonstrating the theoretical possibility that ALP domain walls or composite dark matter objects with similar characteristics~\cite{Col85,Kus01,kimball2018searching} can survive to modern times~\cite{bucher1999dark,avelino2008dynamics,hiramatsu2013axion} and have the characteristics of cold dark matter~\cite{Pos13,Der14,baek2014hidden}.


Since ALPs can interact with atomic spins~\cite{safronova2018search}, the passage of Earth through an ALP domain wall affects atomic spins similarly to a transient magnetic field pulse~\cite{Pos13,kimball2018searching}. Considering a linear coupling between the ALP field gradient $\boldsymbol{\nabla}\axion(\boldsymbol{r},t)$ and atomic spin $\bs{S}$, the interaction Hamiltonian can be written as
\begin{equation}
    {H}\ts{lin} = -(\hbar c)^{3/2} \frac{\cnstRatio}{\SBScale} \frac{\bs{S}}{\lVert S \rVert} \cdot {\bs{\nabla}}\axion(\bs{r},t)~,
\label{Eq:ALP-Hamiltonian-linear}
\end{equation}
where $\hbar$ is the reduced Planck's constant, $c$ is the speed of light, $\bs{r}$ is the position of the spin, $t$ is the time, and $\SBScale/\cnstRatio\equiv\coupling$ is the coupling constant in units of energy described with respect to the symmetry-breaking scale $\SBScale$~\cite{Pus13}, where $\cnstRatio$ is unitless. In most theories, the coupling constants $\coupling$ describing the interaction between Standard Model fermions and the ALP field are proportional to $\SBScale$; however, $\coupling$ can differ between electrons, neutrons, and protons by model-dependent factors that can be significant~\cite{safronova2018search,graham2015experimental}.

In analogy with Eq.~\eqref{Eq:ALP-Hamiltonian-linear}, the Zeeman Hamiltonian describing the interaction of a magnetic field $\boldsymbol{\pseudoMag}$ with an atomic spin $\bs{S}$ can be written as
\begin{equation}
    {H}_\textrm{Z} = -\gamma \bs{S} \cdot \boldsymbol{\pseudoMag} ~,
\label{Eq:ALP-Hamiltonian-Zeeman}
\end{equation}
where $\gamma$ is the gyromagnetic ratio. Since Eqs.~\eqref{Eq:ALP-Hamiltonian-linear}~and~\eqref{Eq:ALP-Hamiltonian-Zeeman} have the same structure, the gradient of the ALP field, even though it couples to the particle spin rather than the magnetic moment, can be treated as a ``pseudo-magnetic field'' as it causes energy shifts of Zeeman sublevels. An important distinction between the ALP-spin interaction [Eq.~\eqref{Eq:ALP-Hamiltonian-linear}] and the Zeeman interaction [Eq.~\eqref{Eq:ALP-Hamiltonian-Zeeman}] is that while $\gamma$ tends to scale inversely with the fermion mass, no such scaling of the ALP-spin interaction is expected~\cite{safronova2018search}.

The amplitude, direction, and duration of the pseudo-magnetic field pulse associated with the transit of the Earth through an ALP domain wall depends on many unknown parameters such as the energy density stored in the ALP field, the coupling constant $\coupling$, the thickness of the domain wall, and the relative velocity $\bs{v}$ between Earth and the domain wall. The dynamical parameters, such as the velocities of the dark matter objects, are expected to randomly vary from encounter-to-encounter. We assume that they are described by the Standard Halo Model for virialized dark matter~\cite{roberts2017search}. Furthermore, the abundance of domain walls in the galaxy is limited by physical constants, $m_\axion$ and $\SBScale$, as these determine the energy contained in the wall and the total energy of all domain walls is constrained by estimates of the local dark matter density~\cite{bovy2012local}. The expected temporal form of the pseudo-magnetic field pulse can depend both on the theoretical model describing the ALP domain wall as well as particular details of the terrestrial encounter such as the orientation of the Earth. The relationships between these parameters and characteristics of the pseudo-magnetic field pulses searched for in our analysis are discussed in Sec.~\ref{app:sensRegion} of the Supplementary Information and Refs.~\cite{Pos13,kimball2018searching,Pus13}.

\begin{figure}
    \centering 
    \includegraphics[width=\oneColWidth]{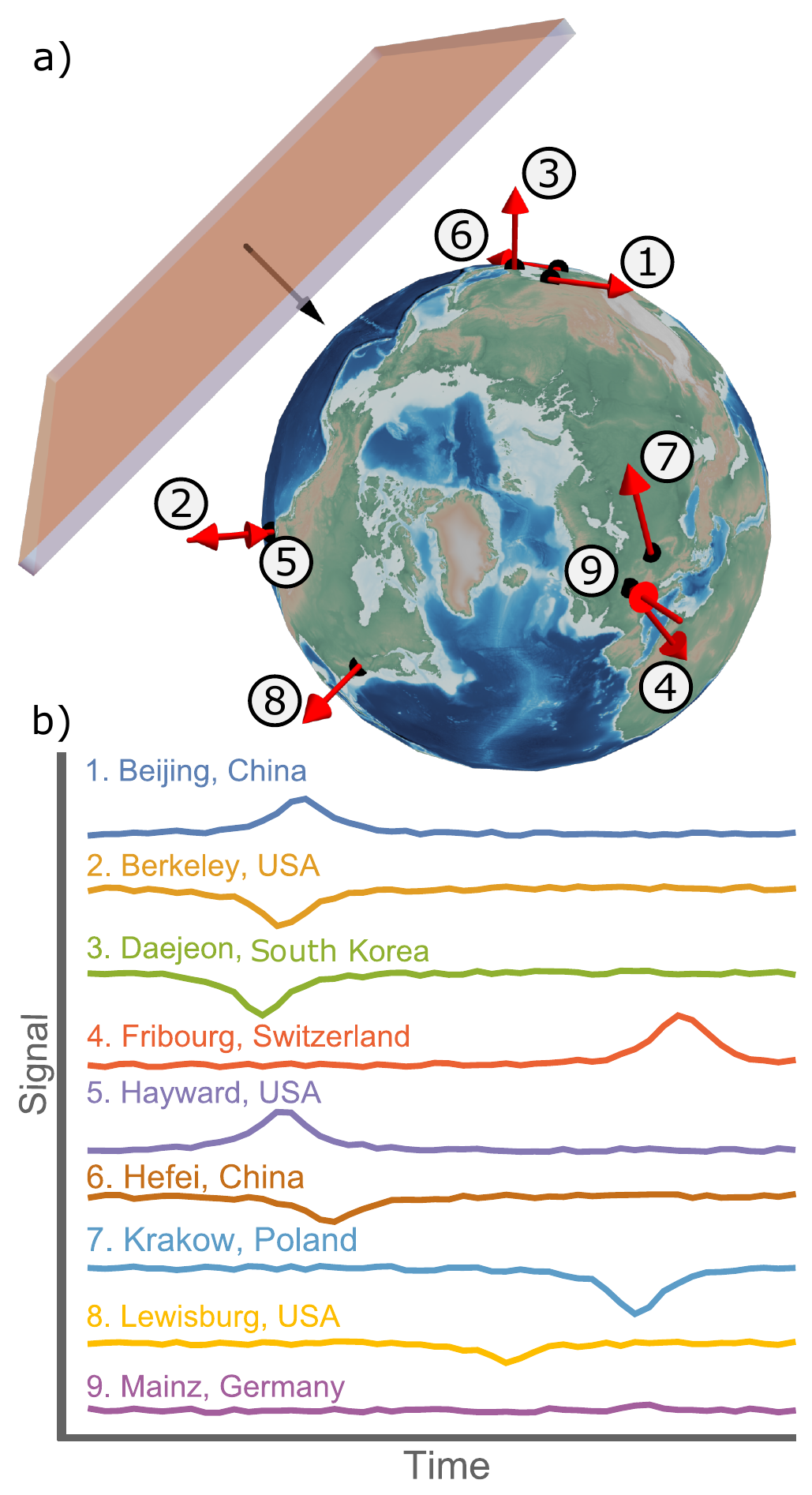} 
    \caption{Visualization of an ALP domain-wall crossing. (a) The image shows the Earth together with the position and sensitive axes of the GNOME magnetometers during Science Run~2.  Position and sensitive axes are show as red arrows. The crossing direction of the domain wall is represented as a black arrow (see Table~\ref{tab:MagInfo}). (b) Simulation of the signals expected to be observed from a domain-wall crossing at the different magnetometers conforming the network. }
    \label{fig:earthWall}
\end{figure}

The Global Network of Optical Magnetometers for Exotic physics searches (GNOME) is a worldwide network searching for correlated signals heralding beyond-the-Standard-Model physics which currently consists of more than a dozen optical atomic magnetometers, with stations (each with a magnetometer and supporting devices) in Europe, North America, Asia, the Middle East, and Australia. A schematic of a domain-wall encounter with the GNOME is shown in Fig.~\ref{fig:earthWall}. The measurements from the magnetometers are recorded with custom data-acquisition systems~\cite{Wlo14}, synchronized to the Global Positioning System (GPS) time, and uploaded to servers located in Mainz, Germany, and Daejeon, South Korea. Descriptions of the operational principles and characteristics of GNOME magnetometers are presented in the Methods and Ref.~\cite{afach2018characterization} (see also Table~\ref{tab:MagInfo}). 

The active field sensor at the heart of every GNOME magnetometer is an optically pumped and probed gas of alkali atoms. Magnetic fields are measured through variations in the Larmor spin precession of the optically polarized atoms. The vapor cells containing the alkali atoms are placed inside multi-layer magnetic shielding systems which reduce background magnetic noise by orders of magnitude~\cite{Bud13} while retaining sensitivity to exotic spin-couplings between ALP dark matter and atomic nuclei.

If the ALP field only couples to electron spins, interaction between the ALP field and the magnetic shield will reduce ALP-induced signal amplitudes in each magnetometer by roughly the magnetic shielding factors of $10^6$--$10^7$ as discussed in Ref.~\cite{Kim16}. Therefore in the present work we only consider interactions between ALP fields and atomic nuclei. Since all GNOME magnetometers presently use atoms whose nuclei have a valence proton, the signal amplitudes measured by the GNOME due to an ALP-spin interaction are proportional to the relative contribution of the proton spin to the nuclear spin (as discussed in Sec.~\ref{app:ProtonCouplings} of the Supplementary Information and Ref.~\cite{Kim15}).  This pattern of signal amplitudes [Eq.~\eqref{Eq:ALP-Hamiltonian-linear}] can be characterized by a pseudo-magnetic field $\pseudoMag_j$ measured with sensor $j$:
\begin{equation}\label{eq:normPMag0}
\pseudoMag_j = \frac{\sigma_{j} \eta_j}{g_{F,j}} \normPMag~,
\end{equation}
where 
\begin{equation}\label{eq:normPMag1}
    \normPMag(\bs{r},t) = \prn{\hbar c}^{3/2} \frac{2\cnstRatio}{\mu_B\SBScale} {\bs{\nabla}}\axion(\bs{r},t) 
\end{equation}
is the normalized pseudo-magnetic field describing the effect of the ALP domain wall on proton spins and $\mu_B$ is the Bohr magneton. The ratio between the Land\'e g-factor and the effective proton spin ($g_{F,j}/\sigma_{j}$) accounts for the specific proton-spin coupling in the respective sensor. This ratio depends on the atomic and nuclear structure as well as details of the magnetometry scheme, see Sec.~\ref{app:ProtonCouplings} of the Supplementary Information. Since each GNOME magnetometer measures the projection of the field along a particular sensitive axis, the factor $\eta_j$ is introduced to account for the directional sensitivity. This factor, given by the cosine of the angle between $\normPMag$ and the sensitive axes, takes on values between $\pm 1$.

In spite of the unknown properties of a particular terrestrial encounter with an ALP domain wall, the GNOME would measure a recognizable global pattern of the associated pseudo-magnetic field pulse amplitudes described by Eq.~\eqref{eq:normPMag0}, as illustrated in Fig.~\ref{fig:earthWall}b. The associated pseudo-magnetic field pulses would point along a common axis, have the same duration, and exhibit a characteristic timing pattern. The data-analysis algorithm used in the present work to search for ALP domain walls is described in the Methods and Ref.~\cite{masia-roig_analysis_2020}. The algorithm searches for a characteristic signal pattern across the GNOME having properties consistent with passage of the Earth through an ALP domain wall. Separate analyses to search for transient oscillatory signals associated with boson stars~\cite{kimball2018searching} and bursts of exotic low-mass fields (ELFs) from cataclysmic astrophysical events~\cite{dailey2020quantum} are presently underway. 

Here we report the results of a dark matter search with the GNOME: a search for transient couplings of atomic spins to macroscopic dark-matter objects, and therefore demonstrate the ability of the GNOME to explore parameter space previously unconstrained by direct laboratory experiments. Searches for macroscopic dark-matter objects based on similar ideas were carried out using atomic clock networks~\cite{Der14, roberts2017search, wcislo2016experimental, roberts2020search}, and there are a number of experimental proposals utilizing other sensor networks~\cite{Sta16, Jac15, Arv16, mcnally2020constraining}. All of these networks are sensitive to bosonic dark matter with a scalar coupling to Standard Model particles~\cite{safronova2018search}. The GNOME is sensitive to a different class of dark matter: bosons with pseudoscalar  couplings to Standard Model particles. Pseudoscalar bosonic dark matter generally produces no observable effects in clock networks~\cite{safronova2018search} but does couple to atomic spins via the interaction described by Eq.~\eqref{Eq:ALP-Hamiltonian-linear}. Thus the GNOME is sensitive to a distinct, so far mostly unconstrained, class of interactions as compared to other sensor networks.

\section*{Search for ALP domain-wall signatures}\label{sec:Results}
There have been four GNOME Science Runs between 2017 and 2020 as discussed in the Methods. Here we analyze the data from Science Run~2, which had comparatively good overall noise characteristics and consistent network operation (as seen in Fig.~\ref{fig:sensOverTime}). Nine magnetometers took part in Science Run~2 that spanned from 29~November 2017 to 22~December 2017. The characteristics of the magnetometers are summarized in Table~\ref{tab:MagInfo}.

Before the data are searched for evidence of domain-wall signatures, they are preprocessed by applying a rolling average, high-pass filters, and notch filters to the raw data. The averaging enhances the signal-to-noise ratio for certain pulse durations, avoids complications arising from different magnetometers having different bandwidths, and reduces the amount of data to be analyzed. The high-pass and notch filters reduce the effects of long-term drifts and noisy frequency bands. We refer to the filtered and rolling-averaged data set as the ``search data.''

The search data are examined for evidence of collective signal patterns corresponding to planes with uniform, non-zero thickness, crossing Earth at constant velocities. The imprinted pattern of amplitudes depends on the domain-wall crossing velocity~\cite{masia-roig_analysis_2020}. We assume that the domain-wall-velocity probability density function follows the Standard Halo Model for virialized dark matter. The signature of a domain-wall crossing the magnetometer network depends on the component of the relative velocity between the domain wall and the Earth that is perpendicular to the domain-wall plane, $\boldsymbol{v}_\bot$. A lattice of points in velocity space is constructed such that the search algorithm covers 97.5\% of the velocity probability density function. The algorithm scans over the velocity lattice and, for every velocity, the data from each magnetometer are appropriately time-shifted so that the signals in different magnetometers from a hypothetical domain-wall crossing with the given velocity occur at the same time. For each velocity and at each measurement time, the amplitudes measured by each magnetometer are fit to the ALP domain-wall crossing model described in Ref.~\cite{masia-roig_analysis_2020}. As a result, estimations for signal magnitude and domain-wall direction, along with associated uncertainties, are obtained for each measurement time and all lattice velocities. The magnitude-to-uncertainty ratio of an event is given by the ratio between the signal magnitude and its associated uncertainty. 

The search algorithm uses two different tests to evaluate if a given event is likely to have been produced by an ALP domain-wall-crossing: a domain-wall model test and a directional-consistency test~\cite{masia-roig_analysis_2020}. The domain-wall model test evaluates whether the event amplitudes measured by the GNOME magnetometers match the signal amplitudes predicted by the ALP domain-wall crossing model, and is quantified by the $p$-value as discussed in the Methods and Ref.~\cite{masia-roig_analysis_2020}. The directional-consistency test checks the agreement between the direction of the scanned velocity and the estimated domain wall direction, and is quantified by the angle between the two directions normalized by the angle between adjacent lattice velocities. The thresholds on these tests are chosen to guarantee an overall detection efficiency $\truePos\geq 95\%$ for the search algorithm, considering both the incomplete velocity lattice coverage and the detection probability (see Fig.~\ref{fig:FN}).

\begin{figure}[ht]
    \centering
    \includegraphics[width=\oneColWidth]{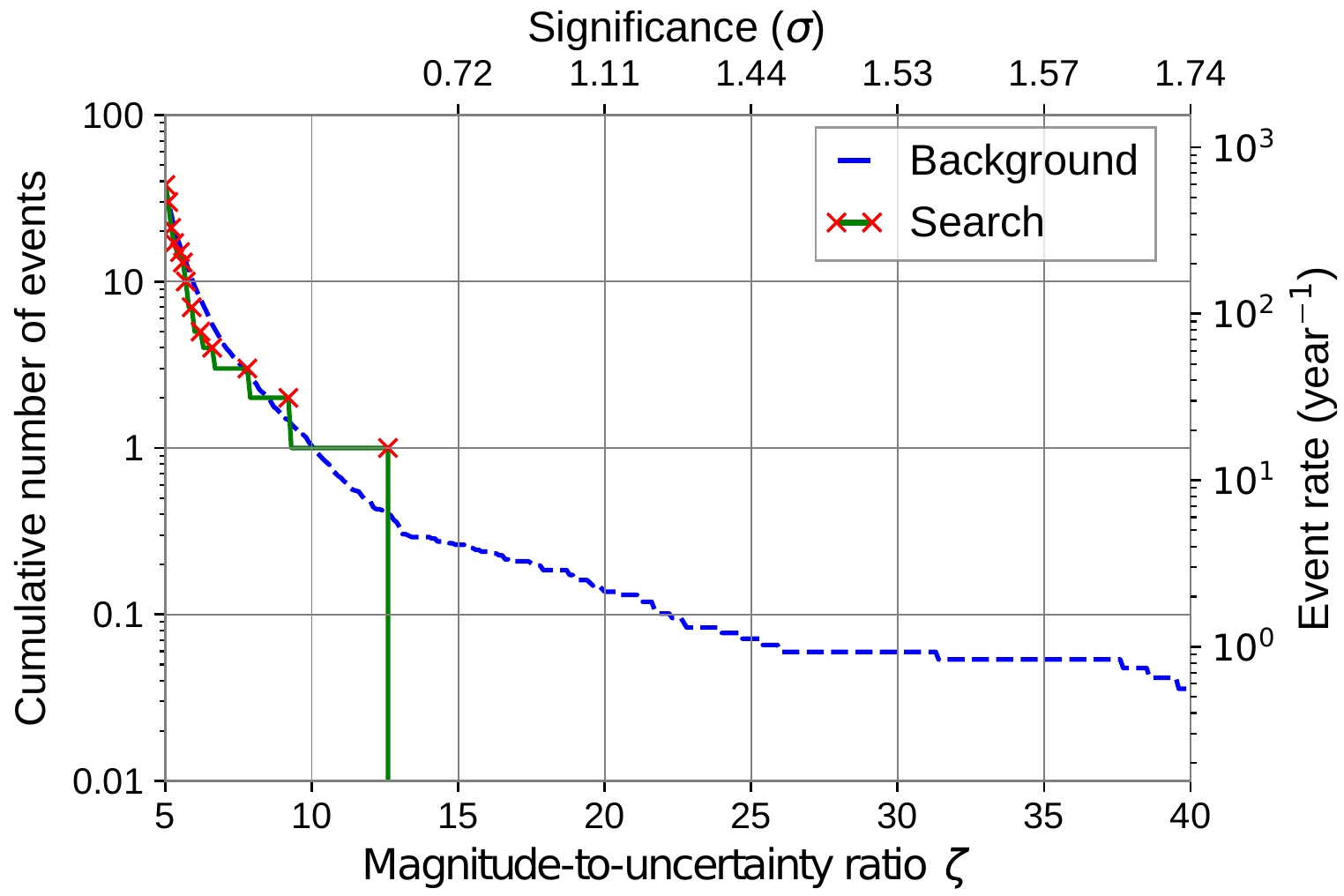} 
    \caption{Significance of the Search events. The blue dashed line represents the cumulative number events expected from the background in the twenty-three days of data from Science Run~2. \BackgroundDuration~years of time-shuffled data are used to evaluate the background. Such duration is an arbitrary choice but is sufficiently long to characterize the background. The number of candidate events measured in the background data is re-scaled to the duration of Science Run~2. The solid green line represents the cumulative number of events measured in Science Run~2. The red crosses indicate at which magnitude-to-uncertainty ratio new events are found in the search data. The upper axis indicates the statistical significance in units of Gaussian standard deviations of finding one event in the search data. The significance is given by the probability of detecting one or more background events at a magnitude-to-uncertainty ratio above that of the candidate event [see Eq.~\eqref{eq:ProbBackgroundEvent}]. The right axis shows the normalized number of events over a period of a year. The event with greatest magnitude-to-uncertainty ratio is found at~\MaxStoN. }
    \label{fig:FP}
\end{figure}

The search data are analyzed for domain-wall encounters using the algorithm presented in Ref.~\cite{masia-roig_analysis_2020}. The cumulative distribution of candidate events as a function of their magnitude-to-uncertainty ratio is shown as a solid green line in Fig.~\ref{fig:FP}. The candidate event in the search data with the largest magnitude-to-uncertainty ratio (= \MaxStoN) had a significance of less than one sigma. Therefore, we find no evidence of an ALP domain-wall crossing during Science Run~2. Rare domain-wall-crossing events producing signals below a magnitude-to-uncertainty ratio of \MaxStoN~are indistinguishable from the background. Therefore, we base constraints on ALP parameters on the absence of any detection above the ``loudest event'' in a manner similar to that described, for example, in Ref.~\cite{LIGO_LoudestEvent}.

\begin{figure}[ht]
    \centering
    \includegraphics[width=\oneColWidth]{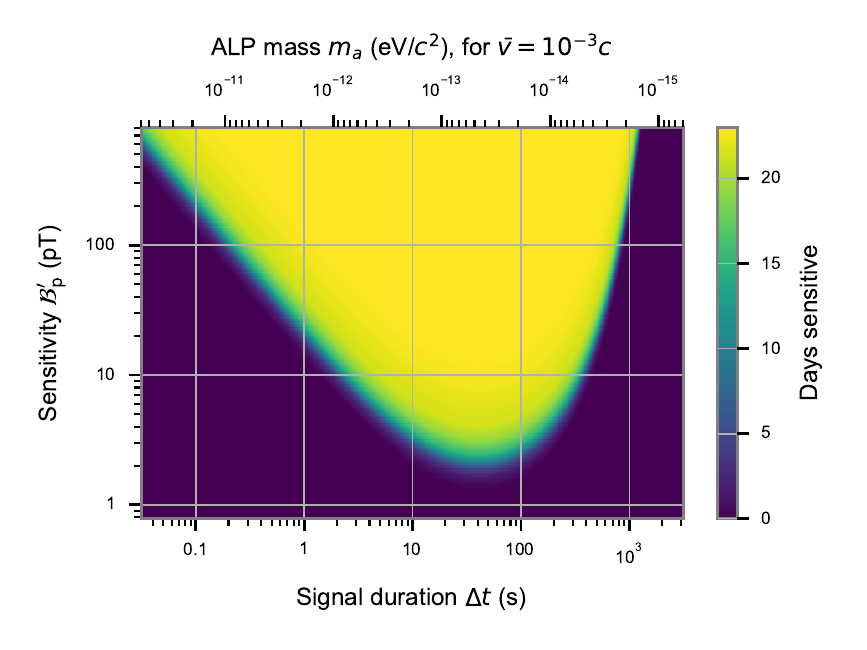}
    \caption{Sensitivity of the GNOME network to domain walls. Amount of time $\effTime$, indicated in color, that the GNOME had a normalized pseudo-magnetic field magnitude sensitivity above $\normPMagSens$ (i.e., the domain wall would induce a magnitude-to-uncertainty ratio of at least one) for domain walls with a given duration $\sigDur$ (defined as the FWHM of a Lorentzian signal) throughout Science Run~2. The upper axis shows the range of ALP masses to which GNOME is sensitive [see Eq.~\eqref{eq:sigDur}]. The characteristic shape of the sensitive region is a result of the filtering and averaging of the raw data as described in the Methods. Averaging reduces the sensitivity of the search data to short pulse durations and high-pass filtering suppresses sensitivity for long $\sigDur$. The GNOME sensitivity varies in time as the number of active GNOME magnetometers recording data and their background noise change. Only the worst-case direction is considered. The plot assumes the parameters of the analysis: 20\,s averaging time, 1.67\,mHz first-order zero-phase Butterworth filter, and 50\,Hz and 60\,Hz zero-phase notch filters with a $Q$-factor of 60.}
    \label{fig:expParamTime}
\end{figure}

In order to evaluate the domain-wall characteristics excluded by this result, the observable domain-wall crossing parameters above~\MaxStoN~magnitude-to-uncertainty ratio during Science Run~2 are determined. The GNOME has nonuniform directional sensitivity~\cite{masia-roig_analysis_2020}; we conservatively estimate the network sensitivity assuming the domain wall comes from the least-sensitive direction. Figure~\ref{fig:expParamTime} shows the active time $\effTime(\sigDur, \normPMagSens)$, i.e., how long the network was sensitive to domain-walls as a function of pseudo-magnetic field-pulse-magnitude sensitivity, $\normPMagSens$, and pulse duration, $\sigDur$. A signal with pseudo-magnetic field magnitude $\normPMag$ produces a magnitude-to-uncertainty ratio of $\SToNSym=\normPMag/\normPMagSens$. The active time, $\effTime(\sigDur, \normPMagSens)$, can be used to constrain ALP domain-wall parameter space as discussed in Sec.~\ref{app:sensRegion} of the Supplementary Information. 

If one assumes a probability distribution for the number of domain-wall encounters, an upper bound on the rate $\RateBound_\totConf$ of such encounters can be calculated with a confidence level $\totConf$. We assume a Poisson probability distribution for the domain-wall crossings. Since the excess number of events in the search data as compared to the background data was not statistically significant, the upper bound on the observable rate is given by the probability of measuring no events during the effective time~\cite{LIGO_LoudestEvent}. Note that since $\effTime$ depends on the parameters of the domain-wall crossing, our constraint on the observed rate depends on the ALP properties. We choose the confidence level to be $\totConf=90\%$.

\section*{Constraints on ALP domain walls}\label{sec:discussion}

The analysis of the GNOME data did not find any statistically significant excess of events above background during Science Run~2 that could point to the existence of ALP domain walls, as seen in Fig.~\ref{fig:FP}. The expected rate of domain-wall encounters, $\evtRate$, depends on the ALP mass, $m_\axion$, the domain-wall energy density in the Milky Way, $\energyDensity$, the typical relative domain-wall speed $\expVel$, and the symmetry breaking scale, $\SBScale$. The region of parameter space to which GNOME is sensitive is defined by the ALP parameters expected to produce signals above~\MaxStoN~magnitude-to-uncertainty ratio with rates $\evtRate \geq \RateBound_{90\%}$ during Science Run~2 (see Fig.~\ref{fig:expParamTime}). Based on the null result of our search, the sensitive region is interpreted as excluded ALP parameter space.

The ALP parameters and the phenomenological parameters describing the ALP domain walls in our galaxy, namely the thickness $\sigWidth$, the surface tension or energy per unit area $\surfTens$, and the average separation $\bar{L}$ can be related through the ALP domain wall model described in Refs.~\cite{Pos13,Pus13}. A full derivation of how observable parameters are related to ALP parameters is given in Sec.~\ref{app:sensRegion} of the Supplementary Information.

\begin{figure*}[ht]
    \centering
    \includegraphics[width=\twoColWidth]{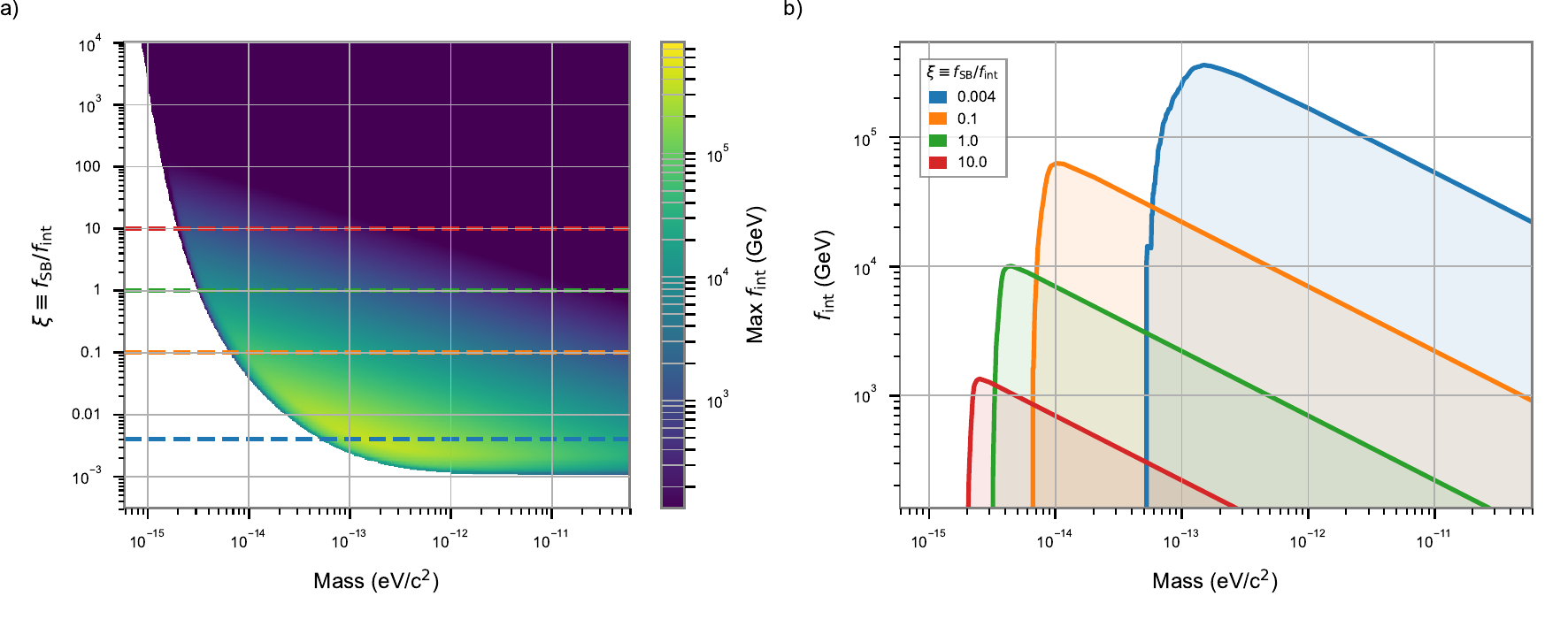}
    \caption{Bounds on the ALP parameter space. The bounds are drawn from the presented analysis of Science Run~2 with 90\% confidence level. The relationship between ALP theory parameters and measured quantities are discussed in Sec.~\ref{app:sensRegion} of the Supplementary Information. (a) In color, upper bound on the the interaction scale for axion-nucleon coupling, $\coupling$, to which the GNOME was sensitive as a function of $m_\axion$ and the ratio between the symmetry-breaking and interaction scales $\cnstRatio\equiv\SBScale/\coupling$. Dashed horizontal lines highlight the cross section used in (b) with the respective color. (b) Cross-sections of the excluded parameter volume in (a) for different ratios $\cnstRatio$. We note that domain walls may not be the only form of dark matter so $\energyDensity < 0.4~{\rm GeV/cm^3}$. If the domain-wall energy density is significantly smaller, this would affect the bounds shown here.}
    \label{fig:physSensRegion}
\end{figure*}

The colored region in Fig.~\ref{fig:physSensRegion}a describes the symmetry breaking scales up to which GNOME was sensitive with 90\% confidence. The parameter space is spanned by ALP mass, maximum symmetry breaking scale, and ratio between symmetry breaking scale and coupling constant. The shape of the sensitive area shown in Fig.~\ref{fig:physSensRegion}a is determined by the event with the largest magnitude-to-uncertainty and the characteristics of the preprocessing applied to the raw data.

Figure~\ref{fig:physSensRegion}b shows various cross-sections of Fig.~\ref{fig:physSensRegion}a for different ratios between the symmetry breaking scale and the coupling constant indicated by the dashed lines. The upper bound of $\SBScale$ that can be observed by the network is shown in Fig.~\ref{fig:physSensRegion}b for different ratios $\cnstRatio\equiv\SBScale/\coupling$. Because $\normPMag \propto m_\axion$ [see Eq.~\eqref{eq:sigMagnitude} in Sec.~\ref{app:sensRegion} of the Supplementary Information], there is a sharp cut-off for low ALP mass where the corresponding field magnitude falls below the network sensitivity. Even though $\normPMag$ increases for large $m_\axion$, the mean rate of domain-wall encounters decreases with increasing mass [see Eqs.~\eqref{eq:DomainScale} and~\eqref{eq:DWrate}]. Correspondingly, the upper limit for the symmetry-breaking scale $\SBScale$ is $\propto 1/\sqrt{m_\axion}$. Given that no events were found, the sensitive region of ALP-domain-wall parameter space during Science Run~2 can be excluded.

Our experiment explores ALP parameter space up to $\coupling \approx 4 \times 10^5~\text{GeV}$ (see Fig.~\ref{fig:physSensRegion}). This goes beyond that excluded by previous direct laboratory experiments searching for ALP-mediated exotic pseudoscalar interactions between protons which have shown that $\coupling \gtrsim 300~{\rm GeV}$ over the ALP mass range probed by the GNOME~\cite{ramsey1979tensor}. Although astrophysical observations suggest that $\coupling \gtrsim 2 \times 10^8~{\rm GeV}$, there are a variety of scenarios in which such astrophysical constraints can be evaded~\cite{derocco2020exploring, bar2020there}. The parameter space for $\coupling$ and $m_\axion$ explored in this search is well outside typical predictions for the quantum chromodynamics (QCD) axion (see, for example, Refs.~\cite{shifman_can_1980, Svr06}). However, for ALPs a vast array of possibilities for the generation of ALP masses and couplings are opened by a variety of beyond-the-standard-model theories, meaning that the values of $\coupling$ and $m_\axion$ explored in our search are theoretically possible (see, for example, the reviews~\cite{marsh2016axion,irastorza2018new} and references therein).

Future work of the GNOME collaboration will focus both on upgrades to our experimental apparatus and new data-analysis strategies. One of our key goals is to improve overall reliability and the duration of continuous operation of GNOME magnetometers. The intermittent operation of some magnetometers due to technical difficulties during Science Runs~1--3 made it difficult to search for signals persisting $\gtrsim$~1~hour.  Additionally, magnetometers varied in their bandwidths and reliability, and stability of their calibration. These challenges were addressed in Science Run~4 through a variety of magnetometer upgrades and instituting daily worldwide test and calibration pulse sequences. However GNOME suffered disruptions due to the worldwide COVID-19 pandemic. We plan to carry out Science Run~5 in 2021 to take full advantage of the improvements. Furthermore, by upgrading to noble-gas-based comagnetometers~\cite{Kor02,Kor05} for future science runs (Advanced GNOME), we expect to significantly improve the sensitivity to ALP domain walls. Additionally, GNOME data can be searched for other signatures of physics beyond the Standard Model such as boson stars~\cite{kimball2018searching}, relaxion halos~\cite{Banerjee2020}, and bursts of exotic low-mass fields from black-hole mergers~\cite{dailey2020quantum}. 

In terms of the data-analysis algorithm used to search for ALP domain walls, recent studies~\cite{stadnik_new_2020} have considered a possible back-action that the Earth may have on a domain wall when certain interactions are significant; namely up-to-quadratic coupling terms between a scalar field and fermions. In contrast to Ref.~\cite{stadnik_new_2020}, the present work analyzes a completely different interaction, namely a linear coupling between a pseudoscalar field and fermion spins, which produces no significant back-action effect. Regardless, it would be worthwhile to consider interactions generating similar back-action effects of the Earth on domain walls and the ALP field in later analysis. Further, in future work we aim to improve the efficiency of the scan over the velocity lattice. The number of points in the velocity lattice to reliably cover a fixed fraction (e.g., 97.5\%) of the ALP velocity probability distribution grows as $\prn{\sigDur}^{-3}$ (where $\sigDur$ is given by Eq.~\eqref{eq:sigDur}). This makes the algorithm computationally intensive. We are investigating a variety of analysis approaches, such as machine-learning-based algorithms, to address these issues.

\section*{Acknowledgments}

The authors are grateful to Chris Pankow, Joshua R. Smith, Jocelyn Read, Menachem Givon, Ron Folman, Wojciech Gawlik, Kathryn Grimm, Grzegorz {\L}ukasiewicz, Peter Fierlinger, Volkmar Schultze, Tilman Sander-Th\"omes, and Holger M\"uller for insightful discussions. 

This work was supported by the U.S. National Science Foundation under grants PHY-1707875, PHY-1707803, PHY-1912465 and PHY-1806672, the Swiss National Science Foundation under grant No. 200021 172686, the German research foundation under grant no. 439720477, the German Federal Ministry of Education and Research (BMBF) within the Quantumtechnologien program (FKZ 13N15064), the European Research Council under the European Union’s Horizon 2020 Research and Innovative Program under Grant agreement No. 695405, the Cluster of Excellence PRISMA+, DFG Reinhart Koselleck Project, Simons Foundation, a Fundamental Physics Innovation Award from the Gordon and Betty Moore Foundation, Heising-Simons Foundation, the National Science Centre of Poland within the OPUS program (Project No. 2015/19/B/ST2/02129), USTC startup funding, and IBS-R017-D1-2021-a00 of the Republic of Korea. The authors acknowledge funding provided by the Institute of Physics Belgrade, trough the grant by the Ministry of Education, Science, and Technological Development of the Republic of Serbia.

\section*{Author contribution statement}
All authors have contributed to the publication, being responsible for the construction and operation of the different magnetometers, building the software infrastructure, assuring the quality of the data being taken, and establishing phenomenological motivation. The data-analysis procedure presented here was led by J.A.S. and H.M.R. with collaboration from other authors. D.F.J.K. coordinated the collaboration between the different teams within GNOME. The manuscript was drafted by H.M.R., J.A.S., A.Wi., and D.F.J.K. It was subject to an internal collaboration-wide review process. All authors reviewed and approved the final version of the manuscript.

\section*{Competing Interests Statement}
The authors declare that they have no competing interests.

\ifuseBibTex{}\else\input{bib_main.bbl}\fi



\section*{Methods}\label{sec:Methods}

The GNOME consists of over a dozen optical atomic magnetometers, each enclosed within a multi-layer magnetic shield, distributed around the world~\cite{Bud13}. GNOME magnetometers are based on a variety of different atomic species, optical transitions, and measurement techniques: some are frequency- or amplitude-modulated nonlinear magneto-optical rotation magnetometers (NMOR)~\cite{budker2002nonlinear, gawlik2006nonlinear}, some are rf-driven optical magnetometers~\cite{afach2018characterization}, while others are spin-exchange-relaxation-free magnetometers (SERF)~\cite{allred2002high}. A detailed description and characterization of six GNOME magnetometers is given in Ref.~\cite{afach2018characterization}. A summary of the properties of the GNOME magnetometers active during Science Run~2 is presented in Table~\ref{tab:MagInfo}. 

\begin{table*}[ht]
\centering
\caption{Characteristics of the magnetometers active during Science Run~2. The station name, location in longitude and latitude, orientation of the sensitive axis, type of magnetometer (NMOR~\cite{budker2002nonlinear, gawlik2006nonlinear}, rf-driven~\cite{afach2018characterization}, or SERF~\cite{allred2002high}), and probed transition are listed. The bandwidth indicates the measured -3\,dB point of the magnetometers' frequency response to oscillating magnetic fields. The calibration error takes into account potential temporal variation of the magnetometers' calibration over the course of Science Run~2, and is estimated based on auxiliary measurements. The rightmost column lists the estimated ratio between the effective proton spin polarization and the Land\'e $g$-factor for the magnetometer, $\sigma_p/g$, which depends on the atomic species and the magnetometry scheme as described in Sec.~\ref{app:ProtonCouplings} of the Supplementary Information. The $\sigma_p/g$ value is used to relate the measured magnetic field to the signal expected from the interaction of an ALP field with proton spins. The indicated uncertainty describes the range of values from different theoretical calculations~\cite{Kim15}.}
\begin{tabular}{ l D{.}{.}{8} D{.}{.}{7} D{.}{.}{0} D{.}{.}{0} l l D{.}{.}{3} D{.}{.}{2} r} 
\hline
\hline
          &                  \multicolumn{2}{c}{Location}                 &           \multicolumn{2}{c}{Orientation}           & \multicolumn{1}{l}{~} & \multicolumn{1}{l}{~} \\
Station   &  \multicolumn{1}{l}{Longitude} &  \multicolumn{1}{l}{Latitude} &   \multicolumn{1}{l}{Az} & \multicolumn{1}{l}{Alt} & \multicolumn{1}{l}{Type} & \multicolumn{1}{l}{Probed transition} &  \multicolumn{1}{l}{Bandwidth} & \multicolumn{1}{l}{Cal. Error } & \multicolumn{1}{l}{~~~~$\sigma_p/g$}\\ 
\hline 
\rule{0ex}{3.6ex} Beijing   & 116.1868\textrm{\textdegree~E} & 40.2457\textrm{\textdegree~N} & +251\textrm{\textdegree} &   0\textrm{\textdegree} & \textrm{NMOR} & \textrm{$^{133}$Cs~D2~} F=4 & 115\,\text{Hz} & 20\% & $-0.39^{+0.19}_{-0.00}$ \\
\rule{0ex}{3.6ex} Berkeley  & 122.2570\textrm{\textdegree~W} & 37.8723\textrm{\textdegree~N} &    0\textrm{\textdegree} & +90\textrm{\textdegree} & \textrm{NMOR} & \textrm{$^{133}$Cs~D2~} F=4 & 7\,\text{Hz} & 40\% & $-0.39^{+0.19}_{-0.00}$\\
\rule{0ex}{3.6ex} Daejeon   & 127.3987\textrm{\textdegree~E} & 36.3909\textrm{\textdegree~N} &    0\textrm{\textdegree} & +90\textrm{\textdegree} & \textrm{NMOR} & \textrm{$^{133}$Cs~D2~} F=4 & 10\,\text{Hz} & 20\% & $-0.39^{+0.19}_{-0.00}$\\ 
\rule{0ex}{3.6ex} Fribourg  &   7.1581\textrm{\textdegree~E} & 46.7930\textrm{\textdegree~N} & +190\textrm{\textdegree} &   0\textrm{\textdegree} & \textrm{rf-driven} & \textrm{$^{133}$Cs~D1~} F=4 & 94\,\text{Hz} & 5\% & $-0.39^{+0.19}_{-0.00}$\\ 
\rule{0ex}{3.6ex} Hayward   & 122.0539\textrm{\textdegree~W} & 37.6564\textrm{\textdegree~N} &    0\textrm{\textdegree} & -90\textrm{\textdegree} & \textrm{NMOR} & \textrm{$^{85}$Rb~D2~} F=3 & 37\,\text{Hz} & 5\% & $-0.36^{+0.05}_{-0.00}$\\ 
\rule{0ex}{3.6ex} Hefei     & 117.2526\textrm{\textdegree~E} & 31.8429\textrm{\textdegree~N} &  +90\textrm{\textdegree} &   0\textrm{\textdegree} & \textrm{SERF} & \textrm{$^{85}$Rb \& $^{87}$Rb~D1~} & 127\,\text{Hz} & 5\% & $-0.38^{+0.05}_{-0.00}$\\ 
\rule{0ex}{3.6ex} Krakow    &  19.9048\textrm{\textdegree~E} & 50.0289\textrm{\textdegree~N} &  +45\textrm{\textdegree} &   0\textrm{\textdegree} & \textrm{NMOR} & \textrm{$^{87}$Rb~D1~} F=2 & 3\,\text{Hz} & 20\% & $~0.50^{+0.00}_{-0.11}$\\ 
\rule{0ex}{3.6ex} Lewisburg &  76.8825\textrm{\textdegree~W} & 40.9557\textrm{\textdegree~N} &    0\textrm{\textdegree} & +90\textrm{\textdegree} & \textrm{SERF} & \textrm{$^{87}$Rb~D2} & 200\,\text{Hz} & 10\% & $~0.70^{+0.00}_{-0.15}$\\  
\rule{0ex}{3.6ex} Mainz     &   8.2354\textrm{\textdegree~E} & 49.9915\textrm{\textdegree~N} &    0\textrm{\textdegree} & -90\textrm{\textdegree} & \textrm{NMOR} & \textrm{$^{87}$Rb~D2~} F=2 & 99\,\text{Hz} & 2\% & $~0.50^{+0.00}_{-0.11}$\\
\hline
\hline
\end{tabular}
\label{tab:MagInfo}
\end{table*}

Each GNOME station is equipped with auxiliary sensors, including accelerometers, gyroscopes, and unshielded magnetometers, to measure local perturbations that could mimic a dark matter signal. Suspicious data are flagged~\cite{afach2018characterization} and discarded during the analysis.

\begin{figure*}
    \centering
    \includegraphics[width=\textwidth]{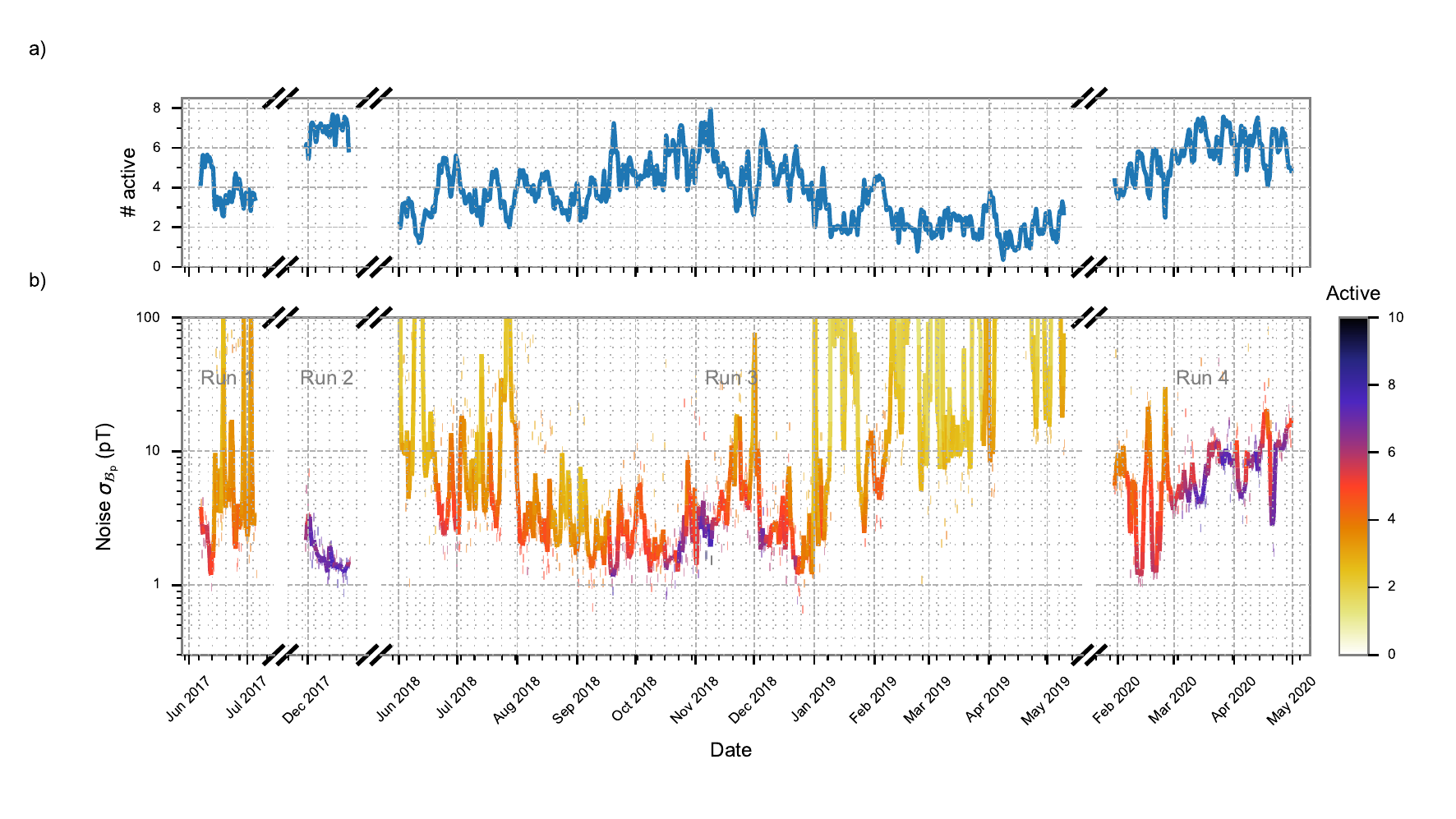}
    \caption{Summary of the GNOME performance during the four Science Runs from 2017 to 2020. The raw magnetometer data are averaged for 20\,s and their standard deviation is calculated over a minimum of one and a maximum of two hours segments depending on the availability of continuous data segments. For each binned point, the combined network noise considering the worse case domain-wall crossing direction is evaluated as defined in Ref.~\cite{masia-roig_analysis_2020}.  (a) One-day rolling average of the number of active sensors. (b)  Multi-colored solid line represents the one-day rolling average of the combined network noise and the multi-colored dashes show the noise of the individual sampled segments. The data are preprocessed with the same filters used for the analysis. The number of magnetometers active is indicated by the color of the line and dashes.} 
    \label{fig:sensOverTime}
\end{figure*}

The number of active GNOME magnetometers during the four Science Runs and the combined network noise as defined in Ref.~\cite{masia-roig_analysis_2020} is shown as a function of time in Fig.~\ref{fig:sensOverTime}. Although Science Run~4 was carried out over a longer period of time than Science Run~2, it featured poorer noise characteristics and consistency of operation compared to Science Run~2. Since many GNOME stations underwent upgrades in 2018 and 2019, further characterization of Science Run~4 data is needed, and results will be presented in future work. The number of active magnetometers during Science Runs~1 and~3 was often less than four, which in insufficient to characterize a domain-wall crossing. We thus present analysis efforts on Science Run~2 data.

Here we provide more details on the analysis procedure. It is composed of three stages to identify events likely to be produced by ALP domain-wall crossings: preprocessing, velocity scanning, and post-selection~\cite{masia-roig_analysis_2020}. First, in the preprocessing stage, a rolling average and filters are applied to the GNOME magnetometer raw data which are originally recorded by the GPS-synchronized data-acquisition system at a rate of 512~samples/s~\cite{Wlo14}. The rolling average is characterized by a 20\,s time constant. Noisy frequency bands are suppressed using a first-order Butterworth high-pass filter at 1.67\,mHz together with the notch filters corresponding to 50\,Hz and 60\,Hz power line frequencies with a quality factor of 60. These filters are applied forward and backward to remove phase effects. This limits the observable pulse properties to a frequency region to which all magnetometers are sensitive. Additionally, it guarantees that the duration of the signal is the same in all sensors. We note that these filter settings may be changed in future analysis. 

The local standard deviation around each point in the magnetometer's data is determined using an iterative process. Outliers are discarded until the standard deviation of the data in the segment converges. The local standard deviation is calculated taking 100 down-sampled points around each data point.

Additionally, auxiliary measurements have shown that the calibration factors used by each magnetometer to convert raw data into magnetic field units experience change over time due to, for example, changes in the environmental conditions. Upper limits on the calibration factor errors due to such drifts over the course of Science Run~2 have been evaluated and are listed in Table~\ref{tab:MagInfo}. Calibration factor errors result in magnetic field measurement errors proportional to the magnetic field $\pseudoMag_j$. The uncertainty resulting from the calibration error is later used to determine agreement with the domain-wall model, but not in the magnitude-to-uncertainty ratio estimate resulting from the model, since the calibration error affects signal and noise in the same way. 

Second, in the velocity-scanning stage, the data from the individual magnetometers are time-shifted according to different relative velocities between Earth and the ALP domain walls. In order to sample 97.5\% of the velocity probability distribution, a scan of the speeds from~53.7 to 770~km/s with directions covering the full $4\pi$ solid angle is chosen, therefore the domain walls can take any orientation with respect to the Earth movement. Note that this distribution considers just the observable perpendicular component of the relative domain-wall velocity and neglects the orbital motion of the Earth around the Sun. For low relative velocities, both the time between signals at different magnetometers as well as the signal duration diverge. So the velocity range is determined by the chosen 97.5\% coverage and the maximum relative speed of domain walls travelling at the galactic escape speed. 

The corresponding time-shifted data along with their local standard deviation estimate are fetched from each magnetometer's rolling averaged full-rate data at a rate of 0.1\,samples/s. This reduces the amount of data to process, while keeping the full timing resolution. 

The step size used in the speed scan is chosen so that a single step in speed corresponds to time-shift differences of less than the down-sampled sampling period. For each speed, a lattice of directions covering the full $4\pi$ solid angle is constructed. The angular difference between adjacent directions is informed by sampling rate and speed~\cite{masia-roig_analysis_2020} such that, as for the speed scan, a single step in direction results in time-shift differences of less than the down-sampled sampling period. With the settings used, the velocity-scanning lattice consists of 1661 points. This number scales with the down-sampled sampling rate cubed.

After time-shifting, the pulses produced by a domain-wall crossing appear simultaneously as if all the magnetometers were placed at the Earth's center. This process results in a time-shifted data set for each lattice velocity on which for each time point a $\chi^2$-minimization is performed to estimate the domain-wall parameters. An ALP domain-wall-crossing direction and magnitude, $\normPMag$, with the corresponding $p$-value quantifying the agreement is obtained. The $p$-value is evaluated as the probability of obtaining the given $\chi^2$ value or higher from the $\chi^2$-minimization. The $p$-value is calculated using the quadrature sum of the standard deviation of the data and the uncertainty due to drifts of the calibration factors. All data points in every time-shifted data set are considered potential events, characterized by time, $p$-value as well as direction and magnitude $\normPMag$ with their associated uncertainties. The magnitude-to-uncertainty ratio of an event $\SToNSym$ is the ratio between this $\normPMag$ and its associated uncertainty.

Third, in the post-selection stage, two tests are carried out to check if a potential event is consistent with an ALP domain-wall crossing. The domain-wall model test evaluates if the observed signal amplitudes are consistent with the expected pattern of a domain-wall crossing from any possible direction. It is quantified by the aforementioned $p$-value. The directional-consistency test is based on the angular difference between the estimated domain-wall crossing direction and the direction of the velocity corresponding to the particular time-shifted data set being analyzed. In a real domain-wall crossing event, these two directions should be aligned. 

To evaluate the consistency of a potential event with a domain-wall crossing, we impose thresholds on the $p$-value and the angular difference normalized with respect to the angular spacing of the lattice of velocity points for that speed. The thresholds are chosen to guarantee a detection probability of 97.5\% with the minimum possible false-positive probability. The false-positive analysis is performed on the background data. The true-positive analysis is performed on test data consisting of background data with randomly inserted domain wall signals as we describe below. 

A single signal pattern may appear as multiple potential events in the analysis, whereas we are seeking to characterize a single underlying domain-wall crossing event. For example, a signal consistent with a domain-wall crossing lasting for multiple sampling periods would appear as multiple potential events in a single time-shifted data set. Furthermore, even if such a signal lasts for only a single sampling period, corresponding potential events appear in different time-shifted data sets. Since it is assumed that domain-wall crossings occur rarely, such clusters of potential events are classified as a single ``event.'' In order to reduce double-counting of these events, conditions are imposed. If potential events passing the thresholds occur at the same time in different time-shifted data sets or are contiguous in time, the potential event with the greatest magnitude-to-uncertainty ratio is classified as the corresponding single event.

In order to evaluate the detection probability of the search algorithm, a well-characterized data set that includes domain-wall-crossing signals with known properties is required. For this purpose, we generate a background data set by randomly time-shuffling the search data so that the relative timing of measurements from different GNOME stations is shifted by amounts so large that no true-positive events could occur. By repeating the process of time shuffling, the length of the background data can be made to far exceed the search data. This method is used to generate background data with noise characteristics closely reproducing those of the search data~\cite{LIGO_GW150914}. A set of pseudo-magnetic field pulses matching the expected amplitude and timing pattern produced by the passages of Earth through ALP domain walls are inserted into the background data to create the test data. 

The true-positive analysis studies the detection probability as a function of the thresholds. Multiple test data sets are created featuring domain-wall-signal patterns with random parameters by inserting Lorentzian-shaped pulses into the background data of the different GNOME magnetometers. The domain-crossing events have magnitudes of $\normPMag$ randomly selected between $0.1$~pT and $10^4$~pT and durations randomly selected between $0.01$~s and $10^3$~s. The distributions of the these randomized parameters are chosen to be flat on a logarithmic scale. Additionally, the signals are inserted at random times with random directions. In order to simulate calibration error effects, the pulse amplitudes inserted in each magnetometer are weighted by a random factor whose range is given in Table~\ref{tab:MagInfo}. The crossing velocity is also randomized within the range covered by the velocity lattice. For each inserted domain-wall-crossing event, the $p$-value, the normalized angular difference, and the magnitude-to-uncertainty ratio are computed.

\begin{figure}[ht]
    \centering
    \includegraphics[width=\columnwidth]{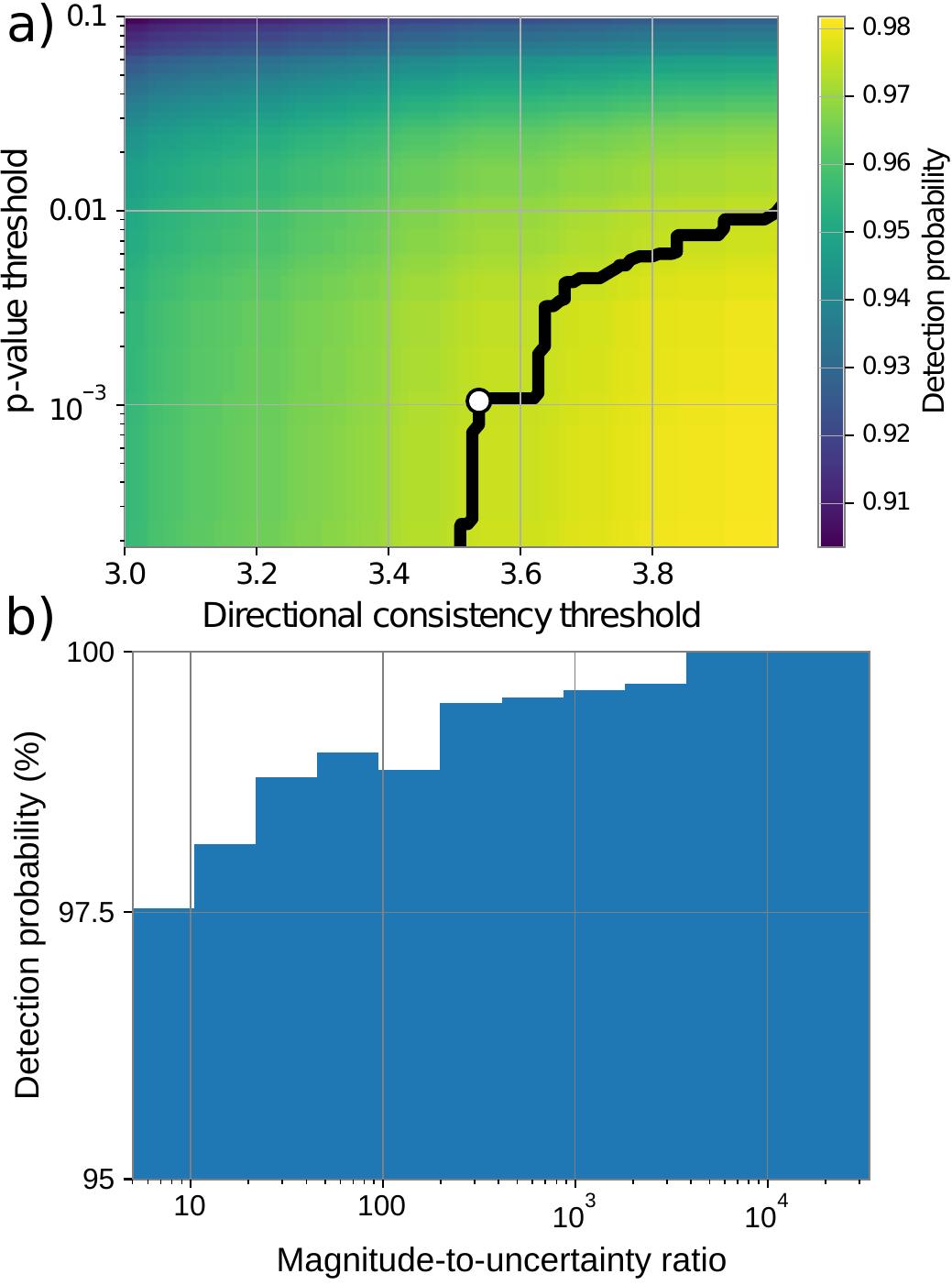} 
    \caption{Summary of the true-positive analysis results. (a) shows the detection probability for domain-wall-crossing event with randomized parameters (as discussed in the text) as a function of $p$-value and directional-consistency thresholds. The inserted events have a magnitude-to-uncertainty ratio between 5 and 10. The black line indicates the combination of parameters corresponding to a 97.5\% detection probability. The white dot indicates the particular thresholds chosen for the analysis. (b) Shows the mean detection probability reached for different magnitude-to-uncertainty ratios for the chosen thresholds.  }
    \label{fig:FN}
\end{figure}

Figure~\ref{fig:FN}a shows the detection probability as a function of the threshold on the lower-limit of the $p$-value and the threshold on the upper-limit of the normalized angular difference. We restrict the analysis in Fig.~\ref{fig:FN}a to events inserted with a magnitude-to-uncertainty ratio between 5 and 10. This enables reliable determination of the true-positive detection probability without significant contamination by false positive events since the background event probability above $\SToNSym=5$ is below 0.01\% in a 10\,s sampling interval. Since the detection probability increases with the signal magnitude, we focus on the events below $\SToNSym=10$. The detection probability is then the number of detected events divided by the number of inserted events. The black line marks the numerically evaluated boundary of the area guaranteeing at least 97.5\% detection. All points along this black line will yield the desired detection probability, so the particular choice is made to minimize the number of candidate events when applying the search algorithm to background data. These values are $\PvalThr$ for the $p$-value threshold and $\AngThr$ for the directional-consistency threshold (represented as a white dot in Fig.~\ref{fig:FN}a). Figure~\ref{fig:FN}b shows that the detection probability is greater than 97.5\% for events featuring a magnitude-to-uncertainty ratio above 5 and guarantees $\truePos \geq 95\%$.This results in an overall detection efficiency $\truePos\geq 95\%$ for the search algorithm, considering both the incomplete velocity lattice coverage and the detection probability.

Since the noise has a nonzero probability of mimicking the signal pattern expected from an ALP domain-wall crossing well enough to pass the $p$-value and directional consistency tests, we perform a false-positive study on background data. The analysis algorithm is applied to $T_b$=\BackgroundDuration~years of time-shuffled data in order to establish the rate of events expected solely from background. Because of the larger amount of background data analyzed, lower rates and larger magnitude-to-uncertainty ratios are accessible as compared to the search data. Based on the false positive study, the probability of finding one or more events in the search data above $\SToNSym$, is~\cite{LIGO_Significance},
\begin{equation}\label{eq:ProbBackgroundEvent}
    P(\geq 1~\text{above}~\SToNSym)=1-\exp{\left(- \frac{\totTime}{T_b} \left[ 1 + n_b(\SToNSym) \right] \right)},
\end{equation}
where $\totTime=23$ days is the duration of Science Run~2 and $n_b(\SToNSym)$ is the number of candidate events found in the background data above $\SToNSym$. The significance is then defined as, $\operatorname{S}=-\sqrt{2} \operatorname{erf}^{-1} \left[1-2 (1-P) \right]$, where $\operatorname{erf}^{-1}$ is the inverse error function. The significance is given in units of the Gaussian standard deviation which corresponds to a one-sided probability of $P$. 

After characterizing the background for Science Run~2, the search data are analyzed. The results are represented as a solid green line in Fig.~\ref{fig:FP}. For $\SToNSym>6$ only a few events were found. The event with largest magnitude-to-uncertainty ratio, $\SToNSym_\text{max}$ was measured at~\MaxStoN~followed by additional events at~$6.2$ and $5.6$. From Eq.~\eqref{eq:ProbBackgroundEvent}, the significance associated with finding one or more events produced by the background featuring at least $\SToNSym_\text{max}$ is lower than one sigma. This null result defines the sensitivity of the search and is used to set constraints on the parameter space describing ALP domain walls.

The observable rate of domain-wall crossings depends on how long GNOME was sensitive to different signal durations and magnitudes. For the evaluation of this effective time, the raw data of each magnetometer are divided into continuous segments between one to two hours duration depending on the availability of the data. The preprocessing steps are applied to each segment. Then the data are binned by taking the average in 20\,s intervals. To estimate the noise in each magnetometer, the standard deviation in each binned segment is calculated to define the covariance matrix $\Sigma_s$. The domain-wall magnitude, crossing with the worse case direction $\hat{\boldsymbol{m}}$, needed to produce $\SToNSym=1$ is calculated as in Ref.~\cite{masia-roig_analysis_2020},
\begin{equation}\label{eq:sensitivity}
    \normPMagSens(\sigDur)=\sqrt{\hat{\boldsymbol{m}}\left( D_{\sigDur}^T \Sigma_s^{-1} D_{\sigDur} \right)^{-1} \hat{\boldsymbol{m}}},
\end{equation}
for each bin. The matrix $D_{\sigDur}$ contains the sensitivity axes of the magnetometers, the factor $\sigma_p / g$, and the effects of the preprocessing as a function of the signal duration as described in Ref.~\cite{masia-roig_analysis_2020}. Such prepocessing effects rely on a Lorentzian-shaped signal and give rise to the characteristic shape of Fig.~\ref{fig:expParamTime}. The effective time, $\effTime$, is defined as the amount of time the network can measure a domain wall with duration $\sigDur$ and magnitude $\normPMagSens$ producing $\SToNSym \geq 1$. Monte Carlo simulations analyzing segments with inserted domain-wall encounters on raw data show a good agreement with the sensitivity estimation in Eq.~\eqref{eq:sensitivity}.

Assuming that the domain-wall encounters follow Poisson statistics, a bound on the observable rate of events above $\SToNSym_\text{max}$ with 90\% confidence is set as~\cite{LIGO_LoudestEvent},
\begin{equation}\label{eq:RateBound}
    \RateBound_{90\%}=\frac{-\log{(0.1)}}{ \truePos~\effTime(\sigDur,\normPMagSens)}.
\end{equation}

The domain-wall thickness is determined by the ALP mass, and is on the order of the ALP reduced Compton wavelength $\lambdabar_\axion$~\cite{Pus13},
\begin{equation}\label{eq:sigWidth}
    \sigWidth \approx 2\sqrt{2} \lambdabar_\axion = 2\sqrt{2} \frac{\hbar}{m_\axion c}\ .
\end{equation}
The constant prefactor of $2\sqrt{2}$ is obtained by approximating the spatial profile of the field-gradient magnitude as a Lorentzian and defining the thickness as full width at half maximum. For a given relative velocity component perpendicular to the domain wall $v_\bot$, the signal duration is
\begin{equation}\label{eq:sigDur}
    \sigDur = \frac{\sigWidth}{v_\bot}\propto m_\axion^{-1}\ .
\end{equation}

We assume that domain walls comprise the dominant component of dark matter. Thus with the energy density $\energyDensity \approx 0.4~{\rm GeV/cm^3} $ in the Milky Way~\cite{bovy2012local}, the energy per unit area (surface tension) in a domain wall, $\surfTens$, determines the average separation between domain walls, $\bar{L}$. The surface tension $\surfTens$ is related to the symmetry breaking scale~\cite{Pos13},
\begin{equation}
    \surfTens = \frac{8}{\hbar^2} m_\axion \SBScale^2\ . 
\end{equation}
The average domain-wall separation is then approximated by
\begin{equation}\label{eq:DomainScale}
    \bar{L} \approx \frac{\surfTens}{\energyDensity} =  \frac{8 }{\hbar^2} \frac{ m_\axion \SBScale^2}{\energyDensity}\ ,
\end{equation}
which results in the average domain-wall encounter rate, 
\begin{equation}\label{eq:DWrate}
    \evtRate = \expVel / \bar{L}\propto \left(m_\axion \SBScale^2\right)^{-1}\ .
\end{equation}
We assume the typical relative domain-wall speed to be equal to the galactic rotation speed of Earth.

The ALP parameter space is constrained by imposing that $\evtRate\geq \RateBound_{90\%}$. The experimental constraint on the coupling constant is written as (see Eq.~\eqref{eq:sensitiveRegion} in Sec.~\ref{app:sensRegion} of the Supplementary Information),
\begin{equation}\label{eq:methods_CouplingLimit}
    \coupling\leq \frac{\hbar}{\cnstRatio}\sqrt{\frac{ \expVel \energyDensity \truePos}{8 m_\axion \log{(0.1)}} \effTime(\sigDur, \normPMagSens)}
\end{equation}
The signal duration can be written in terms of the mass of the hypothetical ALP particle and the specific domain-wall crossing speed, $\sigDur=\frac{2\sqrt{2} \hbar}{\expVel m_\axion c}$. When calculating the constraints on $\coupling$, we fix the domain-wall crossing speed to the typical relative speed from the Standard Halo Model, $\expVel=300$~km s$^{-1}$~\cite{roberts2017search}. In contrast to the signal duration, the pseudo-magnetic field signal depends on all parameters of the ALPs, the mass and the ratio between the coupling and symmetry breaking constants, $\normPMagSens= \frac{4 m_\axion c^{2}  \cnstRatio }{ \mu_B \SToNSym}$. Figure~\ref{fig:physSensRegion} is given by Eq.~\eqref{eq:methods_CouplingLimit} taking $\SToNSym=\MaxStoN$. The shape of the constrained space is given by the fact that $\effTime$ varies depending on the target $m_\axion$ and  $\cnstRatio$. 

\section*{Data availability statement}
The datasets and analysis code used in the current study are available from the corresponding authors on reasonable request. See also collaboration website \url{https://budker.uni-mainz.de/gnome/} where all the data available can be displayed.

\section*{Code availability statement}
The code used in the current study is available under reasonable request.

\ifuseBibTex{}\else\input{bib_methods.bbl}\fi

\ifuseBibTex%
    \bibliographystyle{naturemag}%
    \bibliography{allBib_GNOME-DW-search.bib}%
\fi

\clearpage


\renewcommand{\figurename}{Figure} 
\renewcommand{\tablename}{Table} 
\renewcommand{\theequation}{SI.\arabic{equation}}
\setcounter{equation}{0}
\renewcommand{\thefigure}{SI.\arabic{figure}}
\setcounter{figure}{0}
\renewcommand{\thetable}{SI.\arabic{table}}
\setcounter{table}{0}

\begin{center}
    \Large
    Supplementary Information
\end{center}

\section{Relating measured parameters to ALP domain-wall parameter space}\label{app:sensRegion}

Domain walls form when a field can monotonically vary across vacuum states; two degenerate vacua or possibly the same state if, for example, the field takes values on the 1-sphere.  This is the case for ALPs which arise from the angular part of a complex scalar field~\cite{peccei1977cp}.

The following Lagrangian terms are considered in natural units ($\hbar=c=1$ here and throughout the supplementary information) for a new complex scalar field $\complexScalar$,
\begin{equation}\label{eq:compLag}
\mathcal{L} \supset \lvert \partial_\mu \complexScalar \rvert^2 - \frac{\lambda}{\SNot^{2\NDegVac-4}} \lvert 2^{\NDegVac/2}\complexScalar^\NDegVac + \SNot^\NDegVac \rvert^2\ ,
\end{equation}
where $\lambda$ is a unitless constant and $\SNot$ is a constant with units of energy~\cite{Pos13}. Other references may use a minus sign in front of the $\SNot^N$ term, which results in a similar potential, up to a phase. The end result of Eq.~\eqref{eq:compLag} is that the axion potential will have a maximum at zero, while the minus-convention will have a vacuum at zero.

The axion field is obtained by reparameterizing the complex field $\complexScalar$ in Eq.~\eqref{eq:compLag} in terms of the real field $\HMode$ (Higgs) and $\axion$ (axion), 
\[
\complexScalar=\frac{\HMode}{\sqrt{2}} \exp(i\axion/\SNot)\ .
\]
The second term in Eq.~\eqref{eq:compLag} will break the $U(1)$ symmetry of the complex field into a discrete $\mathbb{Z}_\NDegVac$ symmetry, $\complexScalar\to \exp(2\pi i k/\NDegVac)\complexScalar$ for integer $k$. This corresponds to the axion shift-symmetry, $\axion\to \axion + \frac{2\pi\SNot}{\NDegVac} k$.
The Higgs mode obtains a vacuum expectation value $\HMode\to\SNot$, and the axion field has degenerate vacua or ground energy states at $\axion=\frac{\pi \SNot}{\NDegVac} (2k+1)$ for integer $k$. One can define the symmetry-breaking scale as $\SBScale = \SNot/\NDegVac$. Reparameterizing the complex scalar field in Eq.~\eqref{eq:compLag} and setting the Higgs mode to the vacuum expectation value, the axion Lagrangian is
\begin{equation}\label{eq:axionL2}
\mathcal{L}_\axion = \frac{1}{2} (\partial_\mu \axion)^2 - 2 m_\axion^2 \SBScale^2 \cos^2\left( \frac{\axion}{2\SBScale} \right)\ ,
\end{equation}
where the axion mass is $m_\axion = \NDegVac\SNot\sqrt{2\lambda}$. This can be seen by matching the second-derivative of the cosine term at the minima to a scalar mass term.

For simplicity, a static domain wall in the $yz$-plane separating domains of $-\pi\SBScale$ and $+\pi\SBScale$ is considered. Solving the classical field equations, one finds 
\begin{equation}\label{eq:DWField}
\axion(x) = 2\SBScale\arcsin\left[ \tanh(m_\axion x) \right]\ .
\end{equation}
The gradient of the field is then
\begin{equation}\label{eq:DWGrad}
\frac{da}{dx}(x) = \frac{2\SBScale m_\axion}{\cosh(m_\axion x)}\ .
\end{equation}
This has the full width at half maximum,
\begin{equation}\label{eq:sigFWHM}
\sigWidth = \frac{2\cosh^{-1}(2)}{m_\axion} \approx \frac{2\sqrt{2}}{m_\axion}\ .
\end{equation}
Using the domain-wall solution [Eq.~\eqref{eq:DWField}] and integrating the energy density of the domain wall over $x$ yields the surface tension (energy per unit area)~\cite{Pos13},
\begin{equation}
\surfTens = 8 m_\axion \SBScale^2\ .
\end{equation}

Interactions observable by magnetometers involve coupling between the axion field gradient and the axial-vector current of a fermionic field. For a fermion field $\psi$, the interaction is
\begin{align}
\mathcal{L}_\text{int} =& \frac{i(\complexScalar\partial_\mu\complexScalar^*-\partial_\mu\complexScalar\complexScalar^*)}{\SNot\coupling} \bar{\psi}\gamma^\mu\gamma^5\psi \nonumber \\
&\xrightarrow{S\to\SNot} \frac{\partial_\mu\axion}{\coupling} \bar{\psi}\gamma^\mu\gamma^5\psi\ .
\end{align}
The axial-vector current is related to the spin $\boldsymbol{S}$, so that the interaction Hamiltonian becomes
\begin{equation}\label{eq:intHam}
H_\text{int} = \frac{1}{\coupling} \boldsymbol{\nabla}a \cdot \frac{\boldsymbol{S}}{\lVert S \rVert}\ ,
\end{equation}
i.e., for spin-$\sfrac{1}{2}$ particles, $1/\lVert S \rVert=2$. 

Optical magnetometers operate by measuring the change in atomic energy levels between two energy states with magnetic quantum numbers differing by $\Delta m_F$. The sensitive axis of a magnetometer is either defined by a leading magnetic field applied to the atoms, or, in the case of zero-field magnetometers, by the axis orthogonal to the plane defined by the propagation directions of the pump and probe light.  Variations of the magnetic field along the sensitive axis are measured. The spin coupling from Eq.~\eqref{eq:intHam} can induce a similar shift in energy levels to a magnetic field. The maximum energy shift is determined by plugging the largest gradient from Eq.~\eqref{eq:DWGrad} into Eq.~\eqref{eq:intHam},
\begin{equation}\label{eq:axionEShift}
\Delta E = \sum_{i\in\text{e,p,n,\ldots}}\frac{2\eta \sigma_{(i)} \Delta m_F}{\lVert S_{(i)} \rVert} \frac{\SBScale}{\coupling^{(i)}} m_\axion\ ,
\end{equation}
where $i$ labels the species of fermion, $\sigma_{(i)} = \frac{\left< \boldsymbol{S}_{(i)}\cdot \boldsymbol{F}_{(i)} \right>}{\boldsymbol{F}_{(i)}^2}$ is the projected spin coupling, $\eta=\cos\theta$ for the angle between the axion gradient and sensitive axis $\theta$, and $\coupling^{(i)}$ is the interaction coupling to particle $i$. In general, we will combine the $\frac{\sigma_{(i)}}{\lVert S_{(i)} \rVert\coupling^{(i)}}$ terms into an effective ratio $\frac{2\sigma_j}{\coupling}$, where $j$ now labels the magnetometer. Comparing this to the energy shift due to a magnetic field, $\Delta E_B = g_{F,j}\mu_B\Delta m_F\pseudoMag_j$, one obtains a relationship for a normalized pseudo-magnetic field,
\begin{equation}\label{eq:sigMagnitude}
\frac{g_{F,j} \pseudoMag_j}{\sigma_{j} \eta_j} = \frac{4}{\mu_B} m_\axion \cnstRatio \equiv \normPMag\ ,
\end{equation}
for $\cnstRatio\equiv\frac{\SBScale}{\coupling}$, $g_{F,j}$ being the $g$-factor for the magnetometer $j$, and $\lVert S_{(i)} \rVert=1/2$ since we only consider coupling to spin-$\sfrac{1}{2}$ particles. Here, the normalization is such that $\normPMag$ is the same for all magnetometers, though each individual sensor may observe a different pseudo-magnetic field, $\pseudoMag_j$.

There are two factors that must be considered when determining if axion domain walls are observable by the network: the magnitude of the signal $\normPMag$ and the rate of signals. Domain walls are assumed to exist in a static (or virialized) network across the galaxy through which Earth traverses. For a domain-wall velocity $v$, the duration of a signal is $\sigDur = \sigWidth/v$. Filters and bandwidth limitations generally reduce the magnitude by a factor dependent on the signal duration, which affects the sensitivity of the network (see Appendix in Ref.~\cite{masia-roig_analysis_2020}).

Meanwhile, if the domain walls induce a strong enough signal to be observed, but are so infrequent that one is unlikely to be found over the course of a measurement, then the network is effectively insensitive. If the energy density of domain walls across the galaxy is $\energyDensity$, then the average rate of domain walls passing through Earth is given by
\begin{equation}
    \evtRate = \frac{\expVel \energyDensity}{\surfTens} = \frac{\expVel \energyDensity}{8m_\axion\SBScale^2}\ ,
\end{equation}
where $\expVel$ is the typical relative speed.

The physical parameters describing the ALP domain walls ($m_\axion$, $\SBScale$, and $\coupling$) must be related to the parameters observable by the network ($\normPMag$ and $\sigDur$). The energy density $\energyDensity$ and the typical relative speed $\expVel$ are assumed according to the observed dark matter energy density and the galactic rotation velocity, respectively.

In order to determine if a set of physical parameters is observable, the likelihood that no events are found must be constrained. This constraint defines the confidence level of the detection. The probability of observing $k$ events given that one expects to observe $\mu$ events is given by the Poisson probability mass function,
\[
P(k;\mu) = \frac{\mu^k}{k!} e^{-\mu}\ .
\]
However, the network also has some detection efficiency $\truePos < 1$, so there could be multiple domain-wall-crossing events, but no detection. In particular, the chance of missing no events given that there were $k$ events is $(1-\truePos)^k$. For an event rate $\evtRate$ and measurement time $\totTime$, the probability that no events are detected is then
\[
\sum_{k=0}^\infty (1-\truePos)^k \frac{\left( \evtRate\totTime \right)^k}{k!} e^{-\evtRate\totTime} = e^{-\truePos\evtRate\totTime}\ .
\]
A bound on the event rate $\RateBound_\totConf$ at confidence level $\totConf$ is then given by demanding the probability of observing at least one event $1-e^{-\truePos\RateBound_\totConf\totTime} \geq \totConf$, likewise, one would expect to observe event rates
\begin{equation}
    \evtRate \geq \RateBound_\totConf \equiv \frac{-\log\left( 1-\totConf \right)}{\truePos\totTime}\ .
\end{equation}

The physical parameter space of the ALPs is constrained by demanding that $r\geq\RateBound_\totConf$. Similar arguments for defining constraints can be found, e.g., Ref.~\cite{LIGO_LoudestEvent}. The total time that the network is sensitive to the measurable parameters, $\effTime(\sigDur,\normPMagSens)$, may be less than the total measurement time. These parameters are related to the physical parameters via Eq.~\eqref{eq:sigFWHM} and Eq.~\eqref{eq:sigMagnitude}. One finds a sensitivity bound for $\coupling$ in terms of $m_\axion$ and $\cnstRatio$,
\begin{align}\label{eq:sensitiveRegion}
    \coupling &\leq \frac{1}{\cnstRatio} \sqrt{\frac{-\expVel \energyDensity \truePos}{8m_\axion\log\left( 1-\totConf \right)} \effTime\left( \sigDur, \normPMagSens \right)} \\
    \text{for}\quad \sigDur &= \frac{2\sqrt{2}}{\expVel m_\axion} \quad\text{and}\quad \normPMagSens = \frac{4 m_\axion \cnstRatio}{\mu_B \SToNSym}\nonumber\ ,
\end{align}
where $\normPMagSens$ is the sensitivity of the network and $\SToNSym$ is the magnitude-to-noise ratio induced by a signal with magnitude $\normPMag$. The main calculation from the network data needed for this sensitivity bound is $\effTime$. This is calculated by measuring the sensitivity of the network over time for different signal durations $\sigDur$ and integrating over the time during which the network is sensitive to $\normPMag=\SToNSym\normPMagSens$. Finally, if, after analyzing the data, no domain walls are found, Eq.~\eqref{eq:sensitiveRegion} defines an exclusion region.

\section{Conversion between magnetic field units and proton spin coupling}\label{app:ProtonCouplings}

The amplitude of a signal appearing in the magnetic field data from a GNOME magnetometer due to interaction of atomic spins with an ALP field $\axion$ via the linear coupling described by Eq.~\eqref{Eq:ALP-Hamiltonian-linear} varies based on the atomic species. In every GNOME magnetometer, the atomic vapor cell is located within a multi-layer magnetic shield made of mu-metal and, in some cases a ferrite innermost layer. Interactions of the ALP field with electron spins in the magnetic shielding material can generate a compensating magnetic field that could partially cancel the energy shift due to the interaction of the ALP field with atomic electrons in the vapor cell, as discussed in Ref.~\cite{Kim16}. For this reason, GNOME magnetometers are most sensitive to interactions of the ALP field with nuclear spins. 

\subsection{Deriving spin-projection}

All GNOME magnetometers active during Science Run~2 measure spin-dependent interactions of alkali atoms whose nuclei have valence protons. Thus the GNOME is primarily sensitive to spin-dependent interactions of ALP fields with proton spins. Consequently, the expected signal amplitude measured by a GNOME magnetometer due to the pseudo-magnetic field pulse from passage of Earth through an ALP domain wall must be rescaled by the ratio of the proton spin content of the probed ground-state hyperfine level(s) to their gyromagnetic ratio. Some GNOME magnetometers optically pump and probe a single ground-state hyperfine level, while others rely on the technique of spin-exchange relaxation free (SERF) magnetometry in which the spin-exchange collision rate is much faster than the Larmor precession frequency~\cite{allred2002high,kominis2003subfemtotesla,ledbetter2008spin}. For SERF magnetometers a weighted average of the ground-state Zeeman sublevels over both ground-state hyperfine levels is optically pumped and probed. 

\begin{table*}
\caption{Fractional proton spin polarization $\sigma_{p,F}$, Land\'e $g$-factors $g_F$, and their ratios for the ground state hyperfine levels used in GNOME, and the weighted average of these values across both hyperfine levels ($\abrk{\sigma_{p}}\ts{hf}/\abrk{g}\ts{hf}$) applicable to SERF magnetometers in the low-spin-polarization limit. The estimates are based on the single-particle Schmidt model for nuclear spin~\cite{schmidt1937magnetischen} and the Russell-Saunders scheme for the atomic states. Uncertainties in the values for $\sigma_{p,F}$ describe the range of different results from calculations based on the Schmidt model, semi-empirical models~\cite{engel1989spin, flambaum2006dependence}, and large-scale nuclear shell model calculations where available~\cite{engel1995response, toivanen2009large, iachello1991spin}. The uncertainties in $\sigma_{p,F}$ and $\abrk{\sigma_{p}}\ts{hf}$ are one-sided because alternative methods to the Schmidt model generally predict smaller absolute values of the proton spin polarization. See Ref.~\cite{Kim15} for further details.}
\medskip \begin{tabular}{lrrrr} \hline \hline
\rule{0ex}{3.6ex} Atom~(state)\hspace{2ex} & $\sigma_{p,F}$ & $g_F$ & $\sigma_{p,F}/g_F$ & $\abrk{\sigma_{p}}\ts{hf}/\abrk{g}\ts{hf}$ \\
\hline
\rule{0ex}{3.6ex} $^{39}$K~($F=2$)      & $-0.15^{+0.06}_{-0.00}$ & $~0.50$ & $-0.30^{+0.12}_{-0.00}$ & ~\multirow{2}{*}{-0.5$^{+0.2}_{-0.0}$} \\
\rule{0ex}{3.6ex} $^{39}$K~($F=1$)      & $-0.25^{+0.10}_{-0.00}$ & $-0.50$ & $~0.50^{+0.00}_{-0.19}$  \\
\rule{0ex}{3.6ex} $^{85}$Rb~($F=3$)     & $-0.12^{+0.02}_{-0.00}$ & $~0.33$ & $-0.36^{+0.05}_{-0.00}$ & ~\multirow{2}{*}{-0.8$^{+0.1}_{-0.0}$} \\
\rule{0ex}{3.6ex} $^{85}$Rb~($F=2$)     & $-0.17^{+0.02}_{-0.00}$ & $-0.33$ & $~0.50^{+0.00}_{-0.07}$ \\
\rule{0ex}{3.6ex} $^{87}$Rb~($F=2$)     & $~0.25^{+0.00}_{-0.05}$ & $~0.50$ & $~0.50^{+0.00}_{-0.11}$ & ~~\multirow{2}{*}{0.8$^{+0.0}_{-0.2}$} \\
\rule{0ex}{3.6ex} $^{87}$Rb~($F=1$)     & $~0.42^{+0.00}_{-0.09}$ & $-0.50$ & $-0.83^{+0.18}_{-0.00}$ \\
\rule{0ex}{3.6ex} $^{133}$Cs~($F=4$)    & $-0.10^{+0.05}_{-0.00}$ & $~0.25$ & $-0.39^{+0.19}_{-0.00}$ & ~\multirow{2}{*}{-1.2$^{+0.6}_{-0.0}$} \\
\rule{0ex}{3.6ex} $^{133}$Cs~($F=3$)    & $-0.13^{+0.06}_{-0.00}$ & $-0.25$ & $~0.50^{+0.00}_{-0.24}$ \\
\hline \hline
\end{tabular}
\label{Table:spin-content}
\end{table*}

Table~\ref{Table:spin-content} shows the relevant factors needed to convert the magnetic field signal recorded by GNOME magnetometers into the expected pseudo-magnetic field due to interaction of an ALP field with the proton spin.  Detailed calculations are carried out in Ref.~\cite{Kim15}. The relationship of the expectation value for total atomic angular momentum $\abrk{ \mb{F} }$ to the nuclear spin $\abrk{ \mb{I} }$ can be estimated based on the Russell-Saunders {\it{LS}}-coupling scheme:
\begin{align}
\abrk{ \mb{F} } &= \abrk{ \mb{S}_e } + \abrk{ \mb{L} } + \abrk{ \mb{I} }~, \nonumber \\
&= \frac{\abrk{ \mb{S}_e\cdot\mb{F} } }{ F(F+1) } \abrk{ \mb{F} } + \frac{\abrk{ \mb{L}\cdot\mb{F} }}{ F(F+1) } \abrk{ \mb{F} } + \frac{\abrk{ \mb{I}\cdot\mb{F} }}{ F(F+1) } \abrk{ \mb{F} }~,
\end{align}
where $\mb{S}_e$ is the electronic spin and $\mb{L}$ is the orbital angular momentum. GNOME magnetometers pump and probe atomic states with $L=0$, which simplifies the above equation to:
\begin{equation}
\abrk{ \mb{F} } = \frac{\abrk{ \mb{S}_e\cdot\mb{F} } }{ F(F+1) } \abrk{ \mb{F} } + \frac{\abrk{ \mb{I}\cdot\mb{F} }}{ F(F+1) } \abrk{ \mb{F} }~.
\end{equation}
For $L=0$ the projection of $\mb{I}$ on $\mb{F}$ is given by
\begin{equation}
\abrk{ \mb{I}\cdot\mb{F} }  = \frac{1}{2}\sbrk{F(F+1) + I(I+1) - S_e(S_e+1)}~.
\end{equation}
The above relations define the fractional spin polarization of the nucleus relative to the spin polarization of the atom:
\begin{equation}
\sigma_{N,F} \equiv \frac{\abrk{ \mb{I}\cdot\mb{F} }}{F(F+1)}~.
\end{equation}

The next step is to relate $\sigma_{N,F}$ to the spin polarization of the valence proton $\sigma_{p,F}$ for a particular $F$. As discussed in Ref.~\cite{Kim15}, a reasonable estimate for K, Rb, and Cs nuclei can be obtained from the nuclear shell model by assuming that the nuclear spin $\mb{I}$ is due to the orbital motion and intrinsic spin of only the valence nucleon and that the spin and orbital angular momenta of all other nucleons sum to zero. This is the assumption of the Schmidt or single-particle model~\cite{schmidt1937magnetischen}.  In the Schmidt model, the nuclear spin $\mb{I}$ is generated by a combination of the valence nucleon spin ($\mb{S}_p$) and the valence nucleon orbital angular momentum $\bs{\ell}$, so that we have
\begin{align}
\sigma_{p,F} & = \frac{\abrk{ \mb{S}_{p} \cdot \mb{I} }}{I(I+1)}\sigma_{N,F}~,  \nonumber \\
& = \frac{S_{p}(S_{p}+1) + I(I+1) - \ell(\ell+1)}{2I(I+1)} \sigma_{N,F}~, 
\nonumber  \\ 
& = \sigma_p \sigma_{N,F}~, \label{Eq:Schmidt-model-chiN-2}
\end{align}
where it is assumed that the valence nucleon is in a well-defined state of $\ell$ and $S_{p}$, and $\sigma_p$ is defined to be the fractional proton spin polarization for a given nucleus~\cite{Kim15}.

For comparison between GNOME magnetometers using different atomic species, it is essential to evaluate the uncertainty in the estimate of $\sigma_{p,F}$ based on the Schmidt model. To estimate this uncertainty, we compare calculations of $\sigma_{p,F}$ based on the Schmidt model to the results of the semi-empirical calculations described in Refs.~\cite{engel1989spin, flambaum2006dependence} and to the results of detailed nuclear shell-model calculations where available~\cite{engel1995response, toivanen2009large, iachello1991spin}. Conservatively, we assign the uncertainty in $\sigma_{p,F}$ to be given by the full range (maximum to minimum) of the values of $\sigma_{p,F}$ calculated by these various methods. It turns out that in each considered case, the estimate based on the Schmidt model gives the largest value of $\left|\sigma_{p,F}\right|$, causing the theoretical uncertainties in estimates of $\sigma_{p,F}$ to be one-sided as shown in Table~\ref{Table:spin-content}.  Further details are discussed in Ref.~\cite{Kim15}.

\subsection{SERF magnetometers}

SERF magnetometers operate in a regime where the Larmor frequency is small compared to the spin-exchange rate, so that the rapid spin-exchange locks together the expectation values of the angular momentum projection $\abrk{M_F}$ in both ground-state hyperfine levels of the alkali atom. Because the Land\'{e} $g$-factors $g_F$ in the two ground-state hyperfine levels have nearly equal magnitudes but opposite signs, the magnitude of the effective Land\'{e} $g$-factor in a SERF magnetometer, $\abrk{g}\ts{hf}$, is reduced compared to that in optical atomic magnetometers where a single ground-state hyperfine level is probed.  

To calculate the effective Land\'{e} $g$-factor averaged over hyperfine levels, $\abrk{g}\ts{hf}$, for a SERF magnetometer, it is instructive to consider the equation describing the magnetic torque on an alkali atom,
\begin{equation}
    g_s \mu_B \mb{B} \times \abrk{\mb{S}_e} \approx \dbydt{\abrk{\mb{F}}}~,
    \label{eq:magnetic-torque}
\end{equation}
where $g_s \approx 2$ is the electron $g$-factor and we have ignored the contribution of the nuclear magnetic moment. In the SERF regime, where the alkali vapor is in spin-exchange equilibrium, the populations of the Zeeman sublevels correspond to the spin-temperature distribution~\cite{anderson1959n} described by a density matrix in the Zeeman basis given by~\cite{happer1973spin, savukov2005effects}
\begin{equation}
    \bs{\rho} = Ce^{\bs{\beta}\cdot\mb{F}}~,
\end{equation}
where $C$ is a normalization constant and $\bs{\beta}$ is the spin-temperature vector defined to point in the direction of the spin polarization $P$ with magnitude
\begin{equation}
    \beta = \ln\prn{\frac{1+P}{1-P}}~.
\end{equation}
In the low-spin-polarization limit,
\begin{equation}
    \bs{\rho} \approx C\prn{1 + \bs{\beta}\cdot\mb{F}}~.
\end{equation}
If follows that
\begin{equation}
\abrk{\mb{S}_e} = {\rm{Tr}}\prn{ \bs{\rho} \mb{S}_e } = \frac{1}{3} S_e \prn{S_e+1} \bs{\beta} = \frac{1}{4} \bs{\beta}~,
\label{eq:Se-hf-avg}
\end{equation}
and
\begin{equation}
\abrk{\mb{I}} = {\rm{Tr}}\prn{ \bs{\rho} \mb{I} } = \frac{1}{3} I \prn{I+1} \bs{\beta}~.
\label{eq:I-hf-avg}
\end{equation}
Substituting the above expressions into Eq.~\eqref{eq:magnetic-torque} yields
\begin{equation}
    \abrk{g}\ts{hf} \mu_B \mb{B} \times \bs{\beta} \approx \dbydt{\bs{\beta}}~,
    \label{eq:magnetic-torque-SERF}
\end{equation}
where
\begin{equation}
    \abrk{g}\ts{hf} = \frac{3 g_s}{3 + 4I(I+1)}~.
    \label{eq:SERF-g-factor}
\end{equation}
Equation~\eqref{eq:SERF-g-factor} can be compared to the $g$-factor for a particular alkali ground-state hyperfine level~\cite{budker2008atomic},
\begin{equation}
    g_F = \pm\frac{g_s}{2I+1}~.
    \label{eq:SERF-gF-factor}
\end{equation}

The effective proton spin polarization $\abrk{\sigma_{p}}\ts{hf}$ for SERF magnetometers can also be derived by considering the relevant torque equation
\begin{align}
    \prn{\frac{1}{\coupling} \boldsymbol{\nabla}a} \times \frac{\abrk{\mb{S}_p}}{\lVert S_p \rVert} & = \dbydt{\abrk{\mb{F}}}~, \nonumber  \\
    \prn{\frac{1}{\coupling} \boldsymbol{\nabla}a} \times \frac{\sigma_p}{\lVert S_p \rVert} \abrk{\mb{I}} & = \dbydt{\abrk{\mb{F}}}~,
\end{align}
where we have used the fact that GNOME as configured for Science Run~2 is sensitive to the coupling of the ALP field to proton spins. In the low-spin-polarization limit, based on Eqs.~\eqref{eq:Se-hf-avg} and~\eqref{eq:I-hf-avg},
\begin{equation}
    \prn{\frac{1}{\coupling} \boldsymbol{\nabla}a} \times \frac{\abrk{\sigma_p}\ts{hf}}{\lVert S_p \rVert} \bs{\beta} = \dbydt{\bs{\beta}}~,
\end{equation}
where
\begin{equation}
    \abrk{\sigma_p}\ts{hf} = \sigma_p \frac{4 I (I+1)}{3 + 4I(I+1)}~.
    \label{eq:SERF-proton-spin-polarization}
\end{equation}

Table~\ref{Table:spin-content} shows the ratio between the effective proton spin polarization averaged over both ground-state hyperfine levels, $\abrk{\sigma_p}\ts{hf}$, to the effective Land\'e $g$-factor, $\abrk{g}\ts{hf}$, in the low-spin-polarization regime. Note that the magnitudes of $\abrk{\sigma_{p}}\ts{hf}/\abrk{g}\ts{hf}$ are in general similar or slightly larger than the magnitudes of $\sigma_{p,F}/g_F$ for a single hyperfine level.

To determine the actual values of $\abrk{\sigma_{p}}\ts{hf}/\abrk{g}\ts{hf}$ for the SERF magnetometers used in GNOME's Science Run~2, more detailed considerations are required. 

The values for the ratio of the effective proton spin polarization to the effective Land\'e $g$-factors for each GNOME magnetometer active during Science Run~2 are given in Table~\ref{tab:MagInfo}.

\subsubsection{Hefei magnetometer}

The Hefei GNOME station employs a SERF magnetometer in a closed-loop, single-beam configuration, where the laser light is resonant with the Rb D1 line (pressure-broadened by 600~torr of nitrogen gas to a linewidth of $\sim 10$~GHz). The Hefei SERF magnetometer operates in the low-spin-polarization mode. The vapor cell contains $^{39}$K, $^{85}$Rb, and $^{87}$Rb atoms in natural abundance, so spin-exchange collisions average over both ground-state hyperfine levels of all three species. Taking into account the relative abundances of the different atomic species at the cell temperature of $\approx 150^\circ$C ($\approx 9\%$~$^{39}$K, $\approx 65.5\%$~$^{85}$Rb, $\approx 25.5\%$~$^{87}$Rb), we find that $\abrk{g}\ts{hf} \approx 0.193$, $\abrk{\sigma_{p}}\ts{hf} = -0.073^{+0.010}_{-0.000}$, and thus $\abrk{\sigma_{p}}\ts{hf}/\abrk{g}\ts{hf} = -0.38^{+0.05}_{-0.00}$ for the Hefei magnetometer. 

\subsubsection{Lewisburg magnetometer}

The Lewisburg GNOME station employs a SERF magnetometer in a closed-loop, two-beam configuration. The vapor cell contains only $^{87}$Rb atoms. The Lewisburg SERF magnetometer operates with a spin-polarization $P \approx 0.5$, outside the low-spin-polarization regime. Discussions of the high-polarization regime are given in Refs.~\cite{savukov2005effects,appelt1998theory}. For a nucleus with $I=3/2$, the effective Land\'e $g$-factor is given by
\begin{equation}
    \abrk{g}\ts{hf} = g_s \frac{1+P^2}{6+2P^2}~,
    \label{eq:g-HiPol}
\end{equation}
and the effective proton spin polarization is given by
\begin{equation}
    \abrk{\sigma_{p}}\ts{hf} = \sigma_p \frac{5+P^2}{6+2P^2}~.
    \label{eq:p-spin-HiPol}
\end{equation}
Based on Eqs.~\eqref{eq:g-HiPol} and \eqref{eq:p-spin-HiPol}, we find that for the Lewisburg magnetometer $\abrk{\sigma_{p}}\ts{hf}/\abrk{g}\ts{hf} = 0.70^{+0.00}_{-0.15}$.

\ifuseBibTex{}\else\input{bib_SI.bbl}\fi

\end{document}

%% file: authorList.tex
\author{Samer Afach}
\affiliation{Helmholtz-Institut Mainz, GSI Helmholtzzentrum f{\"u}r Schwerionenforschung, 64291 Darmstadt, Germany}
\affiliation{Johannes Gutenberg-Universit\"at Mainz, 55128 Mainz, Germany}

\author{Ben C. Buchler}
\affiliation{Centre for Quantum Computation and Communication Technology, Research School of Physics, The Australian National University, Acton 2601, Australia}

\author{Dmitry Budker}
\affiliation{Helmholtz-Institut Mainz, GSI Helmholtzzentrum f{\"u}r Schwerionenforschung, 64291 Darmstadt, Germany}
\affiliation{Johannes Gutenberg-Universit\"at Mainz, 55128 Mainz, Germany}
\affiliation{Department of Physics, University of California at Berkeley, Berkeley, California 94720-7300, USA}

\author{Conner Dailey}
\altaffiliation[Currently: ]{University of Waterloo, Department of Physics and Astronomy, N2L 3G1,Ontario, CA}
\affiliation{Department of Physics, University of Nevada, Reno, Nevada 89557, USA}

\author{Andrei Derevianko}
\affiliation{Department of Physics, University of Nevada, Reno, Nevada 89557, USA}

\author{Vincent Dumont}
\affiliation{Computational Research Division, Lawrence Berkeley National Laboratory, Berkeley, CA 94720, USA}

\author{Nataniel L. Figueroa}
\affiliation{Helmholtz-Institut Mainz, GSI Helmholtzzentrum f{\"u}r Schwerionenforschung, 64291 Darmstadt, Germany}
\affiliation{Johannes Gutenberg-Universit\"at Mainz, 55128 Mainz, Germany}

\author{Ilja Gerhardt}
\affiliation{Institute for Quantum Science and Technology (IQST), 3rd Institute of Physics, and Max Planck Institute for Solid State Research, D-70569 Stuttgart, Germany}

\author{Zoran D. Gruji\'{c}}
\affiliation{Institute of Physic Belgrade, University of Belgrade, 11080 Belgrade, Serbia}
\affiliation{Physics Department, University of Fribourg, Chemin du Mus\'ee 3, CH-1700 Fribourg, Switzerland}

\author{Hong Guo}
\affiliation{State Key Laboratory of Advanced Optical Communication Systems and Networks, Department of Electronics, and Center for Quantum Information Technology, Peking University, Beijing 100871, China}

\author{Chuanpeng Hao}
\affiliation{Department of Precision Machinery and Precision Instrumentation, University of Science and Technology of China, Hefei 230026, P. R. China}

\author{Paul S. Hamilton}
\affiliation{Department of Physics and Astronomy, University of California, Los Angeles, California 90095, USA}

\author{Morgan Hedges}
\affiliation{Centre for Quantum Computation and Communication Technology, Research School of Physics, The Australian National University, Acton 2601, Australia}

\author{Derek F. Jackson Kimball}
\affiliation{Department of Physics, California State University -- East Bay, Hayward, California 94542-3084, USA}

\author{Dongok Kim}
\affiliation{Center for Axion and Precision Physics Research, IBS, Daejeon 34051, Republic of Korea}
\affiliation{Department of Physics, KAIST, Daejeon 34141, Republic of Korea}

\author{Sami Khamis}
\affiliation{Department of Physics and Astronomy, University of California, Los Angeles, California 90095, USA}

\author{Thomas Kornack}
\affiliation{Twinleaf LLC, 300 Deer Creek Drive, Plainsboro, NJ 08536, USA}

\author{Victor Lebedev}
\affiliation{Physics Department, University of Fribourg, Chemin du Mus\'ee 3, CH-1700 Fribourg, Switzerland}

\author{Zheng-Tian Lu}
\affiliation{Hefei National Laboratory for Physical Sciences at the Microscale, University of Science and Technology of China, Hefei 230026, P. R. China}

\author{Hector Masia-Roig}
\email{hemasiar@uni-mainz.de}
\affiliation{Helmholtz-Institut Mainz, GSI Helmholtzzentrum f{\"u}r Schwerionenforschung, 64291 Darmstadt, Germany}
\affiliation{Johannes Gutenberg-Universit\"at Mainz, 55128 Mainz, Germany}

\author{Madeline Monroy}
\affiliation{Department of Physics, University of California at Berkeley, Berkeley, California 94720-7300, USA}
\affiliation{Department of Physics, California State University -- East Bay, Hayward, California 94542-3084, USA}

\author{Mikhail Padniuk}
\affiliation{Institute of Physics, Jagiellonian University in Krakow, prof. Stanis{\l}awa {\L}ojasiewicza 11, 30-348, Krak\'ow, Poland}

\author{Christopher A. Palm}
\affiliation{Department of Physics, California State University -- East Bay, Hayward, California 94542-3084, USA}

\author{Sun Yool Park}
\altaffiliation[Currently: ]{JILA, NIST and University of Colorado, and Department of Physics, University of Colorado, Boulder Colorado 80309-0440, USA}
\affiliation{Department of Physics and Astronomy, Oberlin College, Oberlin, OH 44074, USA}

\author{Karun V. Paul}
\affiliation{Centre for Quantum Computation and Communication Technology, Research School of Physics, The Australian National University, Acton 2601, Australia}

\author{Alexander Penaflor}
\affiliation{Department of Physics, California State University -- East Bay, Hayward, California 94542-3084, USA}

\author{Xiang Peng}
\affiliation{State Key Laboratory of Advanced Optical Communication Systems and Networks, Department of Electronics, and Center for Quantum Information Technology, Peking University, Beijing 100871, China}

\author{Maxim Pospelov}
\affiliation{School of Physics and Astronomy, University of Minnesota, Minneapolis, MN 55455, USA}
\affiliation{William I. Fine Theoretical Physics Institute, School of Physics and Astronomy, University of Minnesota, Minneapolis, MN 55455, USA}

\author{Rayshaun Preston}
\affiliation{Department of Physics, California State University -- East Bay, Hayward, California 94542-3084, USA}

\author{Szymon Pustelny}
\affiliation{Institute of Physics, Jagiellonian University in Krakow, prof. Stanis{\l}awa {\L}ojasiewicza 11, 30-348, Krak\'ow, Poland}

\author{Theo Scholtes}
\affiliation{Physics Department, University of Fribourg, Chemin du Mus\'ee 3, CH-1700 Fribourg, Switzerland}
\affiliation{Leibniz Institute of Photonic Technology, Albert-Einstein-Straße 9, D-07745 Jena, Germany}

\author{Perrin C. Segura}
\altaffiliation[Currently: ]{Department of Physics, Harvard University, Cambridge, MA 02138}
\affiliation{Department of Physics and Astronomy, Oberlin College, Oberlin, OH 44074, USA}

\author{Yannis K. Semertzidis}
\affiliation{Center for Axion and Precision Physics Research, IBS, Daejeon 34051, Republic of Korea}
\affiliation{Department of Physics, KAIST, Daejeon 34141, Republic of Korea}

\author{Dong Sheng}
\affiliation{Department of Precision Machinery and Precision Instrumentation, University of Science and Technology of China, Hefei 230026, P. R. China}

\author{Yun Chang Shin}
\affiliation{Center for Axion and Precision Physics Research, IBS, Daejeon 34051, Republic of Korea}

\author{Joseph A. Smiga}
\email{jsmiga@uni-mainz.de}
\affiliation{Helmholtz-Institut Mainz, GSI Helmholtzzentrum f{\"u}r Schwerionenforschung, 64291 Darmstadt, Germany}
\affiliation{Johannes Gutenberg-Universit\"at Mainz, 55128 Mainz, Germany}

\author{Jason E. Stalnaker}
\affiliation{Department of Physics and Astronomy, Oberlin College, Oberlin, OH 44074, USA}

\author{Ibrahim Sulai}
\affiliation{Department of Physics and Astronomy, One Dent Drive, Bucknell University, Lewisburg, Pennsylvania 17837, USA}

\author{Dhruv Tandon}
\affiliation{Department of Physics and Astronomy, Oberlin College, Oberlin, OH 44074, USA}

\author{Tao Wang}
\affiliation{Department of Physics, Princeton University, Princeton, New Jersey, 08544, USA}

\author{Antoine Weis}
\affiliation{Physics Department, University of Fribourg, Chemin du Mus\'ee 3, CH-1700 Fribourg, Switzerland}

\author{Arne Wickenbrock}
\affiliation{Helmholtz-Institut Mainz, GSI Helmholtzzentrum f{\"u}r Schwerionenforschung, 64291 Darmstadt, Germany}
\affiliation{Johannes Gutenberg-Universit\"at Mainz, 55128 Mainz, Germany}

\author{Tatum Wilson}
\affiliation{Department of Physics, California State University -- East Bay, Hayward, California 94542-3084, USA}

\author{Teng Wu}
\affiliation{State Key Laboratory of Advanced Optical Communication Systems and Networks, Department of Electronics, and Center for Quantum Information Technology, Peking University, Beijing 100871, China}

\author{David Wurm}
\affiliation{Technische Universit\"at M\"unchen, 85748 Garching, Germany}

\author{Wei Xiao}
\affiliation{State Key Laboratory of Advanced Optical Communication Systems and Networks, Department of Electronics, and Center for Quantum Information Technology, Peking University, Beijing 100871, China}

\author{Yucheng Yang}
\affiliation{State Key Laboratory of Advanced Optical Communication Systems and Networks, Department of Electronics, and Center for Quantum Information Technology, Peking University, Beijing 100871, China}

\author{Dongrui Yu}
\affiliation{State Key Laboratory of Advanced Optical Communication Systems and Networks, Department of Electronics, and Center for Quantum Information Technology, Peking University, Beijing 100871, China}

\author{Jianwei Zhang}
\affiliation{State Key Laboratory of Advanced Optical Communication Systems and Networks, Department of Electronics, and Center for Quantum Information Technology, Peking University, Beijing 100871, China}

%% file: arxivVer.bbl
\begin{thebibliography}{10}
\expandafter\ifx\csname url\endcsname\relax
  \def\url#1{\texttt{#1}}\fi
\expandafter\ifx\csname urlprefix\endcsname\relax\def\urlprefix{URL }\fi
\providecommand{\bibinfo}[2]{#2}
\providecommand{\eprint}[2][]{\url{#2}}
\setcounter{NAT@ctr}{\value{univBibCtr}}

\bibitem{Ber05}
\bibinfo{author}{Bertone, G.}, \bibinfo{author}{Hooper, D.} \&
  \bibinfo{author}{Silk, J.}
\newblock \bibinfo{title}{Particle dark matter: Evidence, candidates and
  constraints}.
\newblock \emph{\bibinfo{journal}{Phys. Rep.}} \textbf{\bibinfo{volume}{405}},
  \bibinfo{pages}{279--390} (\bibinfo{year}{2005}).

\bibitem{Gor14}
\bibinfo{author}{Gorenstein, P.} \& \bibinfo{author}{Tucker, W.}
\newblock \bibinfo{title}{Astronomical signatures of dark matter}.
\newblock \emph{\bibinfo{journal}{Adv. High Energy Phys.}}
  \textbf{\bibinfo{volume}{2014}}, \bibinfo{pages}{878203}
  (\bibinfo{year}{2014}).

\bibitem{safronova2018search}
\bibinfo{author}{Safronova, M.} \emph{et~al.}
\newblock \bibinfo{title}{Search for new physics with atoms and molecules}.
\newblock \emph{\bibinfo{journal}{Rev. Mod. Phys.}}
  \textbf{\bibinfo{volume}{90}}, \bibinfo{pages}{025008}
  (\bibinfo{year}{2018}).

\bibitem{Pre83}
\bibinfo{author}{Preskill, J.}, \bibinfo{author}{Wise, M.~B.} \&
  \bibinfo{author}{Wilczek, F.}
\newblock \bibinfo{title}{Cosmology of the invisible axion}.
\newblock \emph{\bibinfo{journal}{Phys. Lett. B}}
  \textbf{\bibinfo{volume}{120}}, \bibinfo{pages}{127--132} (\bibinfo{year}{1983}).

\bibitem{graham2015experimental}
\bibinfo{author}{Graham, P.~W.}, \bibinfo{author}{Irastorza, I.~G.},
  \bibinfo{author}{Lamoreaux, S.~K.}, \bibinfo{author}{Lindner, A.} \&
  \bibinfo{author}{van Bibber, K.~A.}
\newblock \bibinfo{title}{Experimental searches for the axion and axion-like
  particles}.
\newblock \emph{\bibinfo{journal}{Annu. Rev. Nucl. Part. Sci.}}
  \textbf{\bibinfo{volume}{65}}, \bibinfo{pages}{485--514}
  (\bibinfo{year}{2015}).

\bibitem{Gra15}
\bibinfo{author}{Graham, P.~W.}, \bibinfo{author}{Kaplan, D.~E.} \&
  \bibinfo{author}{Rajendran, S.}
\newblock \bibinfo{title}{Cosmological relaxation of the electroweak scale}.
\newblock \emph{\bibinfo{journal}{Phys. Rev. Lett.}}
  \textbf{\bibinfo{volume}{115}}, \bibinfo{pages}{221801}
  (\bibinfo{year}{2015}).

\bibitem{co2020predictions}
\bibinfo{author}{Co, R.~T.}, \bibinfo{author}{Hall, L.~J.} \&
  \bibinfo{author}{Harigaya, K.}
\newblock \bibinfo{title}{Predictions for axion couplings from {{ALP}}
  cogenesis}.
\newblock \emph{\bibinfo{journal}{J. High Energ. Phys.}}
  \textbf{\bibinfo{volume}{2021}}, \bibinfo{pages}{172} (\bibinfo{year}{2021}).

\bibitem{Vil85}
\bibinfo{author}{Vilenkin, A.}
\newblock \bibinfo{title}{Cosmic strings and domain walls}.
\newblock \emph{\bibinfo{journal}{Phys. Rep.}} \textbf{\bibinfo{volume}{121}},
  \bibinfo{pages}{263--315} (\bibinfo{year}{1985}).

\bibitem{Pos13}
\bibinfo{author}{Pospelov, M.} \emph{et~al.}
\newblock \bibinfo{title}{Detecting domain walls of axionlike models using
  terrestrial experiments}.
\newblock \emph{\bibinfo{journal}{Phys. Rev. Lett.}}
  \textbf{\bibinfo{volume}{110}}, \bibinfo{pages}{021803}
  (\bibinfo{year}{2013}).

\bibitem{Der14}
\bibinfo{author}{Derevianko, A.} \& \bibinfo{author}{Pospelov, M.}
\newblock \bibinfo{title}{Hunting for topological dark matter with atomic
  clocks}.
\newblock \emph{\bibinfo{journal}{Nat. Phys.}} \textbf{\bibinfo{volume}{10}},
  \bibinfo{pages}{933-–936} (\bibinfo{year}{2014}).

\bibitem{Bra16}
\bibinfo{author}{Braaten, E.}, \bibinfo{author}{Mohapatra, A.} \&
  \bibinfo{author}{Zhang, H.}
\newblock \bibinfo{title}{Dense axion stars}.
\newblock \emph{\bibinfo{journal}{Phys. Rev. Lett.}}
  \textbf{\bibinfo{volume}{117}}, \bibinfo{pages}{121801}
  (\bibinfo{year}{2016}).

\bibitem{kimball2018searching}
\bibinfo{author}{Jackson~Kimball, D.} \emph{et~al.}
\newblock \bibinfo{title}{Searching for axion stars and {Q}-balls with a
  terrestrial magnetometer network}.
\newblock \emph{\bibinfo{journal}{Phys. Rev. D}} \textbf{\bibinfo{volume}{97}},
  \bibinfo{pages}{043002} (\bibinfo{year}{2018}).

\bibitem{sikivie1982axions}
\bibinfo{author}{Sikivie, P.}
\newblock \bibinfo{title}{Axions, domain walls, and the early universe}.
\newblock \emph{\bibinfo{journal}{Phys. Rev. Lett.}}
  \textbf{\bibinfo{volume}{48}}, \bibinfo{pages}{1156--1159} (\bibinfo{year}{1982}).

\bibitem{press1989dynamical}
\bibinfo{author}{Press, W.~H.}, \bibinfo{author}{Ryden, B.~S.} \&
  \bibinfo{author}{Spergel, D.~N.}
\newblock \bibinfo{title}{Dynamical evolution of domain walls in an expanding
  universe}.
\newblock \emph{\bibinfo{journal}{Astrophys. J.}}
  \textbf{\bibinfo{volume}{347}}, \bibinfo{pages}{590--604} (\bibinfo{year}{1989}).

\bibitem{buschmann_early-universe_2020}
\bibinfo{author}{Buschmann, M.}, \bibinfo{author}{Foster, J.~W.} \&
  \bibinfo{author}{Safdi, B.~R.}
\newblock \bibinfo{title}{Early-{{Universe Simulations}} of the {{Cosmological
  Axion}}}.
\newblock \emph{\bibinfo{journal}{Phys. Rev. Lett.}}
  \textbf{\bibinfo{volume}{124}}, \bibinfo{pages}{161103}
  (\bibinfo{year}{2020}).

\bibitem{Col85}
\bibinfo{author}{Coleman, S.}
\newblock \bibinfo{title}{{Q}-balls}.
\newblock \emph{\bibinfo{journal}{Nucl. Phys. B}}
  \textbf{\bibinfo{volume}{262}}, \bibinfo{pages}{263--283} (\bibinfo{year}{1985}).

\bibitem{Kus01}
\bibinfo{author}{Kusenko, A.} \& \bibinfo{author}{Steinhardt, P.~J.}
\newblock \bibinfo{title}{{Q}-ball candidates for self-interacting dark
  matter}.
\newblock \emph{\bibinfo{journal}{Phys. Rev. Lett.}}
  \textbf{\bibinfo{volume}{87}}, \bibinfo{pages}{141301}
  (\bibinfo{year}{2001}).

\bibitem{bucher1999dark}
\bibinfo{author}{Bucher, M.} \& \bibinfo{author}{Spergel, D.}
\newblock \bibinfo{title}{Is the dark matter a solid?}
\newblock \emph{\bibinfo{journal}{Phys. Rev. D}} \textbf{\bibinfo{volume}{60}},
  \bibinfo{pages}{043505} (\bibinfo{year}{1999}).

\bibitem{avelino2008dynamics}
\bibinfo{author}{Avelino, P.}, \bibinfo{author}{Martins, C.},
  \bibinfo{author}{Menezes, J.}, \bibinfo{author}{Menezes, R.} \&
  \bibinfo{author}{Oliveira, J.}
\newblock \bibinfo{title}{Dynamics of domain wall networks with junctions}.
\newblock \emph{\bibinfo{journal}{Phys. Rev. D}} \textbf{\bibinfo{volume}{78}},
  \bibinfo{pages}{103508} (\bibinfo{year}{2008}).

\bibitem{hiramatsu2013axion}
\bibinfo{author}{Hiramatsu, T.}, \bibinfo{author}{Kawasaki, M.},
  \bibinfo{author}{Saikawa, K.} \& \bibinfo{author}{Sekiguchi, T.}
\newblock \bibinfo{title}{Axion cosmology with long-lived domain walls}.
\newblock \emph{\bibinfo{journal}{J. Cosmol. Astropart. Phys.}}
  \textbf{\bibinfo{volume}{2013}}, \bibinfo{pages}{001} (\bibinfo{year}{2013}).

\bibitem{baek2014hidden}
\bibinfo{author}{Baek, S.}, \bibinfo{author}{Ko, P.} \& \bibinfo{author}{Park,
  W.-I.}
\newblock \bibinfo{title}{Hidden sector monopole, vector dark matter and dark
  radiation with {Higgs} portal}.
\newblock \emph{\bibinfo{journal}{J. Cosmol. Astropart. Phys.}}
  \textbf{\bibinfo{volume}{2014}}, \bibinfo{pages}{067} (\bibinfo{year}{2014}).

\bibitem{Pus13}
\bibinfo{author}{Pustelny, S.} \emph{et~al.}
\newblock \bibinfo{title}{The global network of optical magnetometers for
  exotic physics ({GNOME}): A novel scheme to search for physics beyond the
  {Standard Model}}.
\newblock \emph{\bibinfo{journal}{Ann. Phys. (Berl.)}}
  \textbf{\bibinfo{volume}{525}}, \bibinfo{pages}{659--670} (\bibinfo{year}{2013}).


\bibitem{roberts2017search}
\bibinfo{author}{Roberts, B.~M.} \emph{et~al.}
\newblock \bibinfo{title}{Search for domain wall dark matter with atomic clocks
  on board global positioning system satellites}.
\newblock \emph{\bibinfo{journal}{Nat. Commun.}} \textbf{\bibinfo{volume}{8}},
  \bibinfo{pages}{1195} (\bibinfo{year}{2017}).

\bibitem{bovy2012local}
\bibinfo{author}{Bovy, J.} \& \bibinfo{author}{Tremaine, S.}
\newblock \bibinfo{title}{On the local dark matter density}.
\newblock \emph{\bibinfo{journal}{Astrophys. J.}}
  \textbf{\bibinfo{volume}{756}}, \bibinfo{pages}{89} (\bibinfo{year}{2012}).

\bibitem{Wlo14}
\bibinfo{author}{W{\l}odarczyk, P.}, \bibinfo{author}{Pustelny, S.},
  \bibinfo{author}{Budker, D.} \& \bibinfo{author}{Lipi{\'n}ski, M.}
\newblock \bibinfo{title}{Multi-channel data acquisition system with absolute
  time synchronization}.
\newblock \emph{\bibinfo{journal}{Nucl. Instr. Meth. Phys. Res. A}}
  \textbf{\bibinfo{volume}{763}}, \bibinfo{pages}{150--154} (\bibinfo{year}{2014}).

\bibitem{afach2018characterization}
\bibinfo{author}{Afach, S.} \emph{et~al.}
\newblock \bibinfo{title}{Characterization of the global network of optical
  magnetometers to search for exotic physics ({GNOME})}.
\newblock \emph{\bibinfo{journal}{Phys. Dark Universe}}
  \textbf{\bibinfo{volume}{22}}, \bibinfo{pages}{162--180}
  (\bibinfo{year}{2018}).

\bibitem{Bud13}
\bibinfo{author}{Yashchuk, V.~V.}, \bibinfo{author}{Lee, S.-K.} \&
  \bibinfo{author}{Paperno, E.}
\newblock \bibinfo{title}{Magnetic shielding}.
\newblock In \bibinfo{editor}{Budker, D.} \& \bibinfo{editor}{Jackson~Kimball,
  D.~F.} (eds.) \emph{\bibinfo{booktitle}{Optical {Magnetometry}}},
  chap.~\bibinfo{chapter}{12} (\bibinfo{publisher}{Cambridge University Press},
  \bibinfo{year}{2013}).

\bibitem{Kim16}
\bibinfo{author}{Jackson~Kimball, D.~F.} \emph{et~al.}
\newblock \bibinfo{title}{Magnetic shielding and exotic spin-dependent
  interactions}.
\newblock \emph{\bibinfo{journal}{Phys. Rev. D}} \textbf{\bibinfo{volume}{94}},
  \bibinfo{pages}{082005} (\bibinfo{year}{2016}).

\bibitem{Kim15}
\bibinfo{author}{Jackson~Kimball, D.~F.}
\newblock \bibinfo{title}{Nuclear spin content and constraints on exotic
  spin-dependent couplings}.
\newblock \emph{\bibinfo{journal}{New J. Phys.}} \textbf{\bibinfo{volume}{17}},
  \bibinfo{pages}{073008} (\bibinfo{year}{2015}).

\bibitem{masia-roig_analysis_2020}
\bibinfo{author}{Masia-Roig, H.} \emph{et~al.}
\newblock \bibinfo{title}{Analysis method for detecting topological defect dark
  matter with a global magnetometer network}.
\newblock \emph{\bibinfo{journal}{Phys. Dark Universe}}
  \textbf{\bibinfo{volume}{28}}, \bibinfo{pages}{100494}
  (\bibinfo{year}{2020}).
\newblock

\bibitem{dailey2020quantum}
\bibinfo{author}{Dailey, C.} \emph{et~al.}
\newblock \bibinfo{title}{Quantum sensor networks as exotic field telescopes
  for multi-messenger astronomy}.
\newblock \emph{\bibinfo{journal}{Nat. Astron.}} \textbf{\bibinfo{volume}{5}},
  \bibinfo{pages}{150--158} (\bibinfo{year}{2021}).


\bibitem{wcislo2016experimental}
\bibinfo{author}{Wcis{\l}o, P.} \emph{et~al.}
\newblock \bibinfo{title}{Experimental constraint on dark matter detection with
  optical atomic clocks}.
\newblock \emph{\bibinfo{journal}{Nat. Astron.}} \textbf{\bibinfo{volume}{1}},
  \bibinfo{pages}{0009} (\bibinfo{year}{2017}).

\bibitem{roberts2020search}
\bibinfo{author}{Roberts, B.~M.} \emph{et~al.}
\newblock \bibinfo{title}{Search for transient variations of the fine structure
  constant and dark matter using fiber-linked optical atomic clocks}.
\newblock \emph{\bibinfo{journal}{New J. Phys.}} \textbf{\bibinfo{volume}{22}},
  \bibinfo{pages}{093010} (\bibinfo{year}{2020}).

\bibitem{Sta16}
\bibinfo{author}{Stadnik, Y.~V.} \& \bibinfo{author}{Flambaum, V.~V.}
\newblock \bibinfo{title}{Enhanced effects of variation of the fundamental
  constants in laser interferometers and application to dark-matter detection}.
\newblock \emph{\bibinfo{journal}{Phys. Rev. A}} \textbf{\bibinfo{volume}{93}},
  \bibinfo{pages}{063630} (\bibinfo{year}{2016}).

\bibitem{Jac15}
\bibinfo{author}{Jacobs, D.~M.}, \bibinfo{author}{Weltman, A.} \&
  \bibinfo{author}{Starkman, G.~D.}
\newblock \bibinfo{title}{Resonant bar detector constraints on macro dark
  matter}.
\newblock \emph{\bibinfo{journal}{Phys. Rev. D}} \textbf{\bibinfo{volume}{91}},
  \bibinfo{pages}{115023} (\bibinfo{year}{2015}).

\bibitem{Arv16}
\bibinfo{author}{Arvanitaki, A.}, \bibinfo{author}{Dimopoulos, S.} \&
  \bibinfo{author}{Van~Tilburg, K.}
\newblock \bibinfo{title}{Sound of dark matter: Searching for light scalars
  with resonant-mass detectors}.
\newblock \emph{\bibinfo{journal}{Phys. Rev. Lett.}}
  \textbf{\bibinfo{volume}{116}}, \bibinfo{pages}{031102}
  (\bibinfo{year}{2016}).

\bibitem{mcnally2020constraining}
\bibinfo{author}{McNally, R.~L.} \& \bibinfo{author}{Zelevinsky, T.}
\newblock \bibinfo{title}{Constraining domain wall dark matter with a network
  of superconducting gravimeters and {LIGO}}.
\newblock \emph{\bibinfo{journal}{Eur. Phys. J. D}}
  \textbf{\bibinfo{volume}{74}}, \bibinfo{pages}{61} (\bibinfo{year}{2020}).

\bibitem{LIGO_LoudestEvent}
\bibinfo{author}{Brady, P.~R.}, \bibinfo{author}{Creighton, J. D.~E.} \&
  \bibinfo{author}{Wiseman, A.~G.}
\newblock \bibinfo{title}{Upper limits on gravitational-wave signals based on
  loudest events}.
\newblock \emph{\bibinfo{journal}{Class. Quantum Grav.}}
  \textbf{\bibinfo{volume}{21}}, \bibinfo{pages}{S1775--S1781}
  (\bibinfo{year}{2004}).

\bibitem{ramsey1979tensor}
\bibinfo{author}{Ramsey, N.~F.}
\newblock \bibinfo{title}{The tensor force between two protons at long range}.
\newblock \emph{\bibinfo{journal}{Physica A: Statistical Mechanics and its
  Applications}} \textbf{\bibinfo{volume}{96}}, \bibinfo{pages}{285--289}
  (\bibinfo{year}{1979}).

\bibitem{derocco2020exploring}
\bibinfo{author}{DeRocco, W.}, \bibinfo{author}{Graham, P.~W.} \&
  \bibinfo{author}{Rajendran, S.}
\newblock \bibinfo{title}{Exploring the robustness of stellar cooling
  constraints on light particles}.
\newblock \emph{\bibinfo{journal}{Phys. Rev. D}}
  \textbf{\bibinfo{volume}{102}}, \bibinfo{pages}{075015}
  (\bibinfo{year}{2020}).

\bibitem{bar2020there}
\bibinfo{author}{Bar, N.}, \bibinfo{author}{Blum, K.} \&
  \bibinfo{author}{D’amico, G.}
\newblock \bibinfo{title}{Is there a supernova bound on axions?}
\newblock \emph{\bibinfo{journal}{Phys. Rev. D}}
  \textbf{\bibinfo{volume}{101}}, \bibinfo{pages}{123025}
  (\bibinfo{year}{2020}).

\bibitem{shifman_can_1980}
\bibinfo{author}{Shifman, M.}, \bibinfo{author}{Vainshtein, A.} \&
  \bibinfo{author}{Zakharov, V.}
\newblock \bibinfo{title}{Can confinement ensure natural {{CP}} invariance of
  strong interactions?}
\newblock \emph{\bibinfo{journal}{Nuclear Physics B}}
  \textbf{\bibinfo{volume}{166}}, \bibinfo{pages}{493--506} (\bibinfo{year}{1980}).

\bibitem{Svr06}
\bibinfo{author}{Svrcek, P.} \& \bibinfo{author}{Witten, E.}
\newblock \bibinfo{title}{Axions in string theory}.
\newblock \emph{\bibinfo{journal}{J. High Energy Phys.}}
  \textbf{\bibinfo{volume}{2006}}, \bibinfo{pages}{051}
  (\bibinfo{year}{2006}).

\bibitem{marsh2016axion}
\bibinfo{author}{Marsh, D.~J.}
\newblock \bibinfo{title}{Axion cosmology}.
\newblock \emph{\bibinfo{journal}{Phys. Rep.}} \textbf{\bibinfo{volume}{643}},
  \bibinfo{pages}{1--79} (\bibinfo{year}{2016}).

\bibitem{irastorza2018new}
\bibinfo{author}{Irastorza, I.~G.} \& \bibinfo{author}{Redondo, J.}
\newblock \bibinfo{title}{New experimental approaches in the search for
  axion-like particles}.
\newblock \emph{\bibinfo{journal}{Prog. Part. Nucl. Phys.}}
  \textbf{\bibinfo{volume}{102}}, \bibinfo{pages}{89--159} (\bibinfo{year}{2018}).

\bibitem{Kor02}
\bibinfo{author}{Kornack, T.~W.} \& \bibinfo{author}{Romalis, M.~V.}
\newblock \bibinfo{title}{Dynamics of two overlapping spin ensembles
  interacting by spin exchange}.
\newblock \emph{\bibinfo{journal}{Phys. Rev. Lett.}}
  \textbf{\bibinfo{volume}{89}}, \bibinfo{pages}{253002}
  (\bibinfo{year}{2002}).

\bibitem{Kor05}
\bibinfo{author}{Kornack, T.~W.}, \bibinfo{author}{Ghosh, R.~K.} \&
  \bibinfo{author}{Romalis, M.~V.}
\newblock \bibinfo{title}{Nuclear spin gyroscope based on an atomic
  comagnetometer}.
\newblock \emph{\bibinfo{journal}{Phys. Rev. Lett.}}
  \textbf{\bibinfo{volume}{95}}, \bibinfo{pages}{230801}
  (\bibinfo{year}{2005}).

\bibitem{Banerjee2020}
\bibinfo{author}{Banerjee, A.}, \bibinfo{author}{Budker, D.},
  \bibinfo{author}{Eby, J.}, \bibinfo{author}{Kim, H.} \&
  \bibinfo{author}{Perez, G.}
\newblock \bibinfo{title}{Relaxion stars and their detection via atomic
  physics}.
\newblock \emph{\bibinfo{journal}{Commun. Phys.}} \textbf{\bibinfo{volume}{3}}, \bibinfo{pages}{1}
  (\bibinfo{year}{2020}).

\bibitem{stadnik_new_2020}
\bibinfo{author}{Stadnik, Y.~V.}
\newblock \bibinfo{title}{New bounds on macroscopic scalar-field topological
  defects from nontransient signatures due to environmental dependence and
  spatial variations of the fundamental constants}.
\newblock \emph{\bibinfo{journal}{Phys. Rev. D}}
  \textbf{\bibinfo{volume}{102}}, \bibinfo{pages}{115016}
  (\bibinfo{year}{2020}).

\setcounter{univBibCtr}{\value{NAT@ctr}}
\end{thebibliography}

\begin{thebibliography}{10}
\expandafter\ifx\csname url\endcsname\relax
  \def\url#1{\texttt{#1}}\fi
\expandafter\ifx\csname urlprefix\endcsname\relax\def\urlprefix{URL }\fi
\providecommand{\bibinfo}[2]{#2}
\providecommand{\eprint}[2][]{\url{#2}}
\setcounter{NAT@ctr}{\value{univBibCtr}}

\bibitem{budker2002nonlinear}
\bibinfo{author}{Budker, D.}, \bibinfo{author}{Kimball, D.},
  \bibinfo{author}{Yashchuk, V.} \& \bibinfo{author}{Zolotorev, M.}
\newblock \bibinfo{title}{Nonlinear magneto-optical rotation with
  frequency-modulated light}.
\newblock \emph{\bibinfo{journal}{Phys. Rev. A}} \textbf{\bibinfo{volume}{65}},
  \bibinfo{pages}{055403} (\bibinfo{year}{2002}).

\bibitem{gawlik2006nonlinear}
\bibinfo{author}{Gawlik, W.} \emph{et~al.}
\newblock \bibinfo{title}{Nonlinear magneto-optical rotation with amplitude
  modulated light}.
\newblock \emph{\bibinfo{journal}{Appl. Phys. Lett.}}
  \textbf{\bibinfo{volume}{88}}, \bibinfo{pages}{131108}
  (\bibinfo{year}{2006}).

\bibitem{allred2002high}
\bibinfo{author}{Allred, J.}, \bibinfo{author}{Lyman, R.},
  \bibinfo{author}{Kornack, T.} \& \bibinfo{author}{Romalis, M.~V.}
\newblock \bibinfo{title}{High-sensitivity atomic magnetometer unaffected by
  spin-exchange relaxation}.
\newblock \emph{\bibinfo{journal}{Phys. Rev. Lett.}}
  \textbf{\bibinfo{volume}{89}}, \bibinfo{pages}{130801}
  (\bibinfo{year}{2002}).

\bibitem{LIGO_GW150914}
\bibinfo{author}{Abbott, B.~P.} \emph{et~al.}
\newblock \bibinfo{title}{{GW150914}: First results from the search for binary
  black hole coalescence with {Advanced LIGO}}.
\newblock \emph{\bibinfo{journal}{Phys. Rev. D}} \textbf{\bibinfo{volume}{93}},
  \bibinfo{pages}{122003} (\bibinfo{year}{2016}).

\bibitem{LIGO_Significance}
\bibinfo{author}{Usman, S.~A.} \emph{et~al.}
\newblock \bibinfo{title}{The {PyCBC} search for gravitational waves from
  compact binary coalescence}.
\newblock \emph{\bibinfo{journal}{Class. Quantum Grav.}}
  \textbf{\bibinfo{volume}{33}}, \bibinfo{pages}{215004}
  (\bibinfo{year}{2016}).

\setcounter{univBibCtr}{\value{NAT@ctr}}
\end{thebibliography}

\begin{thebibliography}{10}
\expandafter\ifx\csname url\endcsname\relax
  \def\url#1{\texttt{#1}}\fi
\expandafter\ifx\csname urlprefix\endcsname\relax\def\urlprefix{URL }\fi
\providecommand{\bibinfo}[2]{#2}
\providecommand{\eprint}[2][]{\url{#2}}
\setcounter{NAT@ctr}{\value{univBibCtr}}

\bibitem{peccei1977cp}
\bibinfo{author}{Peccei, R.~D.} \& \bibinfo{author}{Quinn, H.~R.}
\newblock \bibinfo{title}{{CP} conservation in the presence of instantons}.
\newblock \emph{\bibinfo{journal}{Phys. Rev. Lett.}}
  \textbf{\bibinfo{volume}{38}}, \bibinfo{pages}{1440--1443}
  (\bibinfo{year}{1977}).

\bibitem{kominis2003subfemtotesla}
\bibinfo{author}{Kominis, I.}, \bibinfo{author}{Kornack, T.},
  \bibinfo{author}{Allred, J.} \& \bibinfo{author}{Romalis, M.~V.}
\newblock \bibinfo{title}{A subfemtotesla multichannel atomic magnetometer}.
\newblock \emph{\bibinfo{journal}{Nature}} \textbf{\bibinfo{volume}{422}},
  \bibinfo{pages}{596--599} (\bibinfo{year}{2003}).

\bibitem{ledbetter2008spin}
\bibinfo{author}{Ledbetter, M.}, \bibinfo{author}{Savukov, I.},
  \bibinfo{author}{Acosta, V.}, \bibinfo{author}{Budker, D.} \&
  \bibinfo{author}{Romalis, M.}
\newblock \bibinfo{title}{Spin-exchange-relaxation-free magnetometry with {Cs}
  vapor}.
\newblock \emph{\bibinfo{journal}{Phys. Rev. A}} \textbf{\bibinfo{volume}{77}},
  \bibinfo{pages}{033408} (\bibinfo{year}{2008}).

\bibitem{schmidt1937magnetischen}
\bibinfo{author}{Schmidt, T.}
\newblock \bibinfo{title}{{\"U}ber die magnetischen momente der atomkerne}.
\newblock \emph{\bibinfo{journal}{Z. Phys.}} \textbf{\bibinfo{volume}{106}},
  \bibinfo{pages}{358--361} (\bibinfo{year}{1937}).

\bibitem{engel1989spin}
\bibinfo{author}{Engel, J.} \& \bibinfo{author}{Vogel, P.}
\newblock \bibinfo{title}{Spin-dependent cross sections of weakly interacting
  massive particles on nuclei}.
\newblock \emph{\bibinfo{journal}{Phys. Rev. D}} \textbf{\bibinfo{volume}{40}},
  \bibinfo{pages}{3132--3135} (\bibinfo{year}{1989}).

\bibitem{flambaum2006dependence}
\bibinfo{author}{Flambaum, V.} \& \bibinfo{author}{Tedesco, A.}
\newblock \bibinfo{title}{Dependence of nuclear magnetic moments on quark
  masses and limits on temporal variation of fundamental constants from atomic
  clock experiments}.
\newblock \emph{\bibinfo{journal}{Phys. Rev. C}} \textbf{\bibinfo{volume}{73}},
  \bibinfo{pages}{055501} (\bibinfo{year}{2006}).

\bibitem{engel1995response}
\bibinfo{author}{Engel, J.}, \bibinfo{author}{Ressell, M.},
  \bibinfo{author}{Towner, I.} \& \bibinfo{author}{Ormand, W.}
\newblock \bibinfo{title}{Response of mica to weakly interacting massive
  particles}.
\newblock \emph{\bibinfo{journal}{Phys. Rev. C}} \textbf{\bibinfo{volume}{52}},
  \bibinfo{pages}{2216--2221} (\bibinfo{year}{1995}).

\bibitem{toivanen2009large}
\bibinfo{author}{Toivanen, P.}, \bibinfo{author}{Kortelainen, M.},
  \bibinfo{author}{Suhonen, J.} \& \bibinfo{author}{Toivanen, J.}
\newblock \bibinfo{title}{Large-scale shell-model calculations of elastic and
  inelastic scattering rates of lightest supersymmetric particles ({LSP}) on
  $^{127}${I}, $^{129}${Xe}, $^{131}${Xe}, and $^{133}${Cs} nuclei}.
\newblock \emph{\bibinfo{journal}{Phys. Rev. C}} \textbf{\bibinfo{volume}{79}},
  \bibinfo{pages}{044302} (\bibinfo{year}{2009}).

\bibitem{iachello1991spin}
\bibinfo{author}{Iachello, F.}, \bibinfo{author}{Krauss, L.~M.} \&
  \bibinfo{author}{Maino, G.}
\newblock \bibinfo{title}{Spin-dependent scattering of weakly interacting
  massive particles in heavy nuclei}.
\newblock \emph{\bibinfo{journal}{Physics Letters B}}
  \textbf{\bibinfo{volume}{254}}, \bibinfo{pages}{220--224}
  (\bibinfo{year}{1991}).

\bibitem{anderson1959n}
\bibinfo{author}{Anderson, L.~W.}, \bibinfo{author}{Pipkin, F.~M.} \&
  \bibinfo{author}{Baird~Jr, J.~C.}
\newblock \bibinfo{title}{{N}$^{14}$ - {N}$^{15}$ hyperfine anomaly}.
\newblock \emph{\bibinfo{journal}{Phys, Rev.}} \textbf{\bibinfo{volume}{116}},
  \bibinfo{pages}{87--98} (\bibinfo{year}{1959}).

\bibitem{happer1973spin}
\bibinfo{author}{Happer, W.} \& \bibinfo{author}{Tang, H.}
\newblock \bibinfo{title}{Spin-exchange shift and narrowing of magnetic
  resonance lines in optically pumped alkali vapors}.
\newblock \emph{\bibinfo{journal}{Phys. Rev. Lett.}}
  \textbf{\bibinfo{volume}{31}}, \bibinfo{pages}{273--276} (\bibinfo{year}{1973}).

\bibitem{savukov2005effects}
\bibinfo{author}{Savukov, I.} \& \bibinfo{author}{Romalis, M.}
\newblock \bibinfo{title}{Effects of spin-exchange collisions in a high-density
  alkali-metal vapor in low magnetic fields}.
\newblock \emph{\bibinfo{journal}{Phys. Rev. A}} \textbf{\bibinfo{volume}{71}},
  \bibinfo{pages}{023405} (\bibinfo{year}{2005}).

\bibitem{budker2008atomic}
\bibinfo{author}{Budker, D.}, \bibinfo{author}{Kimball, D.~F.} \&
  \bibinfo{author}{DeMille, D.~P.}
\newblock \emph{\bibinfo{title}{Atomic physics: an exploration through problems
  and solutions}} (\bibinfo{publisher}{Oxford University Press, USA},
  \bibinfo{year}{2008}).

\bibitem{appelt1998theory}
\bibinfo{author}{Appelt, S.} \emph{et~al.}
\newblock \bibinfo{title}{Theory of spin-exchange optical pumping of $^{3}${He} and
  $^{129}${Xe}}.
\newblock \emph{\bibinfo{journal}{Phys. Rev. A}} \textbf{\bibinfo{volume}{58}},
  \bibinfo{pages}{1412--1439} (\bibinfo{year}{1998}).

\setcounter{univBibCtr}{\value{NAT@ctr}}
\end{thebibliography}
